\documentclass{emulateapj}

\slugcomment{Draft}

\shorttitle{$X_{\rm CO}$ Gradient in NGC 628}
\shortauthors{Blanc et al.}

\begin{document}

\title{The VIRUS-P Exploration of Nearby Galaxies (VENGA): The
  $X_{\rm CO}$ Gradient in NGC 628}

\author{Guillermo A. Blanc\altaffilmark{1}, 
Andreas Schruba\altaffilmark{2}, 
Neal J. Evans II\altaffilmark{3}, 
\mbox{Shardha Jogee \altaffilmark{3},} 
Alberto Bolatto\altaffilmark{4}, 
Adam K. Leroy\altaffilmark{5}, 
Mimi Song \altaffilmark{3}, 
\mbox{Remco C. E. van den Bosch \altaffilmark{6},} 
Niv Drory \altaffilmark{7}, 
Maximilian Fabricius \altaffilmark{8}, 
David Fisher\altaffilmark{4},
Karl Gebhardt \altaffilmark{3}, 
Amanda Heiderman \altaffilmark{3}, 
Irina Marinova \altaffilmark{3},
Stuart Vogel \altaffilmark{4}, 
Tim Weinzirl \altaffilmark{3}}

\altaffiltext{1}{Observatories of the Carnegie Institution for Science, Pasadena, CA, USA}
\altaffiltext{2}{Astronomy Department, California Institute of Technology, Pasadena, CA, USA}
\altaffiltext{3}{Astronomy Department, The University of Texas at Austin, Austin, TX, USA}
\altaffiltext{4}{Department of Astronomy, University of Maryland, College Park, MD, USA}
\altaffiltext{5}{National Radio Astronomy Observatory, Charlotsville, VA, USA}
\altaffiltext{6}{Max Planck Institute for Astronomy, Heidelberg, Germany}
\altaffiltext{7}{Instituto de Astronomia, Universidad Nacional Autonoma de Mexico, Mexico DF, Mexico}
\altaffiltext{8}{Max Planck Institute for Extraterrestrial Physics, Garching, Germany}
\altaffiltext{9}{Shanghai Astronomical Observatory, Shanghai, China}
\altaffiltext{10}{Astronomy Department, University of Washington, Seattle, WA}

\begin{abstract}

We measure the radial profile of the $^{12}$CO(1-0) to H$_2$ conversion
factor ($X_{\rm CO}$) in NGC 628. The H$\alpha$ emission from
the VENGA integral field spectroscopy is used to map the star formation rate surface density ($\Sigma_{SFR}$).
We estimate the molecular gas surface density ($\Sigma_{H2}$) from $\Sigma_{SFR}$ by inverting the
molecular star formation law (SFL), and compare it to the CO intensity
to measure $X_{\rm CO}$. We study the impact of systematic uncertainties by changing the
slope of the SFL, using different SFR tracers (H$\alpha$
vs. far-UV plus 24$\mu$m), and CO maps from different
telescopes (single-dish and interferometers). The observed $X_{\rm
  CO}$ profile is robust against these systematics, drops by a
factor of 2 from \mbox{$R\sim7$ kpc} to the center of the galaxy,
and is well fit by a gradient \mbox{$\Delta {\rm  log}(X_{\rm
    CO})=0.06\pm0.02$ dex kpc$^{-1}$}. 
We study how changes in $X_{\rm CO}$ follow changes in metallicity,
gas density, and ionization parameter. Theoretical
models show that the gradient in $X_{\rm CO}$ can be explained by
a combination of decreasing metallicity, and decreasing $\Sigma_{H2}$
with radius. Photoelectric heating from the local UV radiation field
appears to contribute to the decrease of $X_{CO}$ in higher density
regions. Our results show that galactic environment plays an
important role at setting the physical conditions in star forming
regions, in particular the chemistry of carbon in molecular
complexes, and the radiative transfer of CO emission. We caution
against adopting a single $X_{\rm CO}$ value when large
changes in gas surface density or metallicity are present.

\end{abstract}


\section{Introduction}

Measuring and studying molecular gas in galaxies is
fundamental to understand star formation, and the physical processes
setting the balance between the different phases of the interstellar
medium (ISM). Molecular hydrogen (H$_2$) amounts for the bulk of the
mass in molecules in the universe, but its observable transitions are
rarely excited at the typically cold temperatures ($\sim$10 K) of the
gas inside giant molecular clouds (GMCs). To overcome
this observational difficulty, the second most abundant molecule in GMCs, the carbon
monoxide molecule $^{12}$C$^{16}$O (hereafter CO), is typically used as a proxy for estimating the total
mass in H$_2$. Rotational CO transitions, observed at millimeter (mm)
wavelengths, can be easily excited under the typical density and
temperature conditions in GMCs, and are therefore bright enough to
be detectable in single molecular clouds in the Milky Way (MW), galaxies
within the Local Group, across the disks of nearby galaxies out to
distances of $\sim10$ Mpc, and even out to high redshifts by integrating the
emission over whole galaxies.

Using CO emission to estimate the H$_2$ mass requires knowledge of the
CO intensity to H$_2$ column density conversion
factor:

\begin{equation}
X_{\rm CO}=\frac{N(H_2)}{I(CO)},
\end{equation}

\noindent
which for Galactic molecular clouds in the vicinity of the Sun, has
typical values of $2-4\times10^{20}$  cm$^{-1}$(K km
  s$^{-1}$)$^{-1}$ \citep[][and references therein]{kennicutt12}. Or
alternatively the CO luminosity to total mass conversion factor
\mbox{$\alpha_{CO}=M_{H2}/L(CO)$} which unlike $X_{\rm CO}$ includes a
factor of 1.36 for the contribution of Helium to the total
mass\footnote{
We report our results in terms of $X_{\rm CO}$, but $\alpha_{CO}$ can be
easily derived using the following relation: $\alpha_{CO}\;[{\rm    M_{\odot}pc^{-2}(K\;km\;s^{-1})^{-1}}]=X_{\rm CO}\;[{\rm
    cm^{-2}(K\;km\;s^{-1})^{-1}}]/(4.6\times10^{19})$}.

Constraining the appropriate value of this
conversion factor and establishing how
it changes under the different physical conditions present across
different environments inside and
across galaxies is of paramount importance if astrophysical
interpretations regarding the molecular ISM are to be drawn from CO
data. Of particular interest is the metallicity dependance of
$X_{\rm CO}$, as deep observations with current facilities, and the advent
of new and more powerful telescopes like ALMA and CCAT, allow the
detection of CO in dwarf galaxies, the outskirts of the disks of
massive spirals, and high redshift systems, where the
heavy element abundance is expected to be low. 

A series of studies have used different techniques to explore the
possibility of a changing $X_{\rm CO}$ across different types of galaxies
in the nearby universe. These include virial mass measurements of
individual GMCs in the Milky Way, the local group, and nearby spirals
\citep[e.g.][and references therein]{wilson95, blitz07, bolatto08, fukui10}, estimating the molecular gas
mass from dust far-IR emission modeling while constraining the dust-to-gas
ratio and the contribution from atomic hydrogen \citep{israel97,
  leroy11, leroy12},
and using the star formation rate (SFR) under the
assumption of a known molecular gas depletion timescale to estimate
the amount of H$_2$ \citep{schruba12, mcquinn12}. A consistent picture seems to arise from these studies,
in which $X_{\rm CO}$ shows higher values for lower metallicity
systems. The difference can be dramatic for the lowest metallicity
dwarfs in the local universe, where the conversion factor can be 10 to 100 times higher than
in the Milky Way. This increase is most likely driven not only by a decrease in the
carbon and oxygen abundances, but mainly by a drop in the optical
depth within GMCs due to a lower abundance of dust. The latter
translates in the CO/C$^+$ dissociation boundary moving inwards within
these clouds, leaving behind large envelopes of ``CO dark'' molecular
gas \citep[e.g.][]{bolatto99}.

On the other hand, studies of molecular gas in merging and starburst
galaxies (typically ultra-luminous infrared galaxies, ULIRGS) based on
virial mass measurements, dust emission modeling, and column density
estimation from optically thin transitions of CO isotopes, find
$X_{\rm CO}$ values which are factors of a few lower than
the typical MW values in the solar vicinity \citep{wild92, shier94, mauersberger96, solomon97,
  downes98, bryant99, meier10}. The same effect is observed for
ULIRGS and sub-mm galaxies (SMGs) at high redshift ($z>1$) by
\cite{solomon05} and \cite{tacconi08}. This effect is thought to be
caused by the impact of higher gas temperatures and stronger
turbulence on the brightness temperature of the CO line and the escape
probability of CO (1-0) photons. Since the CO(1-0)
transition is typically optically thick, the broadening of the
line-width ($\Delta v$) induced by higher levels of turbulence
in these high density environments promotes the escape of CO(1-0) photons \citep{shetty11b}.
Interestingly, the $X_{\rm CO}$ dependance with metallicity mentioned in
the last paragraph has also been observed in a sample of more
``normal'' star forming galaxies at high redshift by \cite{genzel12}
using a method similar to the one used in this work. 

These observational efforts to measure changes in $X_{\rm CO}$ across
different environments have been accompanied in the last few years by
detailed theoretical modeling attempts to understand how CO
radiative transfer depends on the physical conditions of the ISM.
A series of studies using analytic models, numerical simulations, and
combinations of both, have examined the dependance of $X_{\rm CO}$ with
metallicity, gas temperature, gas dynamics, and the local radiation
field \citep[e.g.][]{krumholz11, shetty11b, narayanan12, feldmann12},
although see early work by \cite{dickman86} and \cite{maloney88}.

Considering the fact that $X_{\rm CO}$ changes from galaxy to galaxy depending
on the average physical conditions of the ISM, it would not be
surprising if it also changes within galaxies, depending on the local
physical conditions present in different environments inside an
individual system. For years, this has been known to be the case in
our own galaxy. The value of $X_{\rm CO}$ has been shown to change as a
function of galactocentric radius in the MW using a series of
different techniques: dust emission modeling \citep{sodroski95},
measurements of gamma-ray emissivity from cosmic-ray gas interactions
\citep{digel96, strong04, abdo10}, and direct virial mass measurements of GMCs
\citep{arimoto96, oka98}. All these studies find a decrease in $X_{\rm CO}$ from the typical
values measured near the solar radius ($2-4\times10^{20}$  cm$^{-1}$(K
km s$^{-1}$)$^{-1}$) towards smaller galactocentric radii, with the
conversion factor reaching typical starburst/merger type values of
$0.1-0.5\times 10^{20}$ cm$^{-1}$(K km s$^{-1}$)$^{-1}$ in the
Galactic Center. It is not clear from current studies if the $X_{\rm CO}$
radial profile in the Milky Way follows a smooth gradient or if it
is fairly constant across the Galactic disk and falls sharply at some
intermediate radius.

Measurements of the spatial distribution of $X_{\rm CO}$ within spiral
galaxies are scarce in the literature, and the subject remains highly
unexplored from an observational perspective. Observations of the optically
thin CO isotopes $^{13}$CO and C$^{18}$O, and dust continuum emission,
have been used to show that $X_{\rm CO}$ is a factor of 2-4 lower than the
typical solar vicinity MW values in the central regions of the nearby
barred spirals \mbox{NGC 6946} and \mbox{Maffei 2} \citep{meier04, meier08}, confirming the results from an
earlier study of \mbox{NGC 253}, \mbox{IC 342}, \mbox{Maffei 2}, 
and \mbox{NGC 6946} by \cite{wall93}. Similar work by
\cite{villa-vilaro08} has shown a decreased $X_{\rm CO}$ factor in the
central kpc of \mbox{NGC 5194} (a.k.a. M51a). Using an independent method
based on estimating the dust mass distribution by conducting radiative transfer
modeling of optical and near-IR images of galaxies, and assuming a
constant dust-to-gas ratio, \cite{regan00} also finds low
$X_{\rm CO}$ values in the central regions ($<1.5$ kpc) of \mbox{NGC 1068}, \mbox{NGC
1530}, \mbox{NGC 2903}, and \mbox{NGC 6946}. These results are consistent
with what is observed in the MW. The radial profile of $X_{\rm
  CO}$ out to large radii in spiral galaxies other than the MW remains
largely unexplored.

Only three published measurements of the $X_{\rm CO}$ radial profile for
spiral galaxies other than the Milky Way are known to the authors. First, the
work of \cite{arimoto96} in \mbox{NGC 5194} (M51a),
who use GMC virial mass estimates to find that $X_{\rm CO}$ follows a
linear gradient inside one effective radius in this galaxy. Their
measurements are based on the data of \cite{adler92} which has a
$\sim$350 pc beam-size, and we do not consider virial masses measured
on such physical scales to be reliable. Second, is the high
resolution (beam-size $\sim$20 pc) study of GMCs in M33 by \cite{rosolowsky03}, who use virial
masses derived from CO line-widths to find a flat $X_{\rm CO}$ distribution
across the galaxy, even in the presence of a 0.8 dex change in
metallicity across their sample. And, finally, the soon to be published
dust modeling study of a sample of nearby spirals by
Karin Sandstrom (private communication), which is discussed in \S4 \citep[see also][]{aniano12}. The
spatial distribution of the conversion factor in external galaxies is
the subject of this paper, in which we study the radial profile of the $X_{\rm CO}$
across the disk of the nearby face-on Sc galaxy \mbox{NGC 628}. 

Our method to measure $X_{\rm CO}$ is based on the correlation
observed over many orders of magnitude between the SFR surface density
and the surface density of molecular gas in the ISM of star forming galaxies. This relation, typically
known as the ``star formation law'' (SFL) or the ``Schimdt-Kennicutt
law'' \citep{schmidt59, kennicutt98}, has been recently constrained in
a spatially resolved manner across the disks of nearby spiral galaxies
\citep{kennicutt07, bigiel08, blanc09, verley10, onodera10, schruba11,
  liu11,
  rahman12}. Although debate persists regarding the actual value of
the SFL slope, which is subject to a series of systematic uncertainties
related to background subtraction, cloud sampling, and fitting methods
\citep[see discussions in][]{blanc09, rahman12, calzetti12}, its
normalization is consistent with a depletion
timescale for molecular gas of $\sim 2$ Gyr at the typical molecular
gas surface densities observed across the disks of nearby spirals
\citep[$\Sigma_{H2}=10-100$ M$_{\odot}$pc$^{-2}$, e.g.][]{leroy08, rahman12}. This
relation can be used to derive the molecular gas surface density from
the observed SFR surface density across the disk of a galaxy,
therefore permitting the measurement of $X_{\rm CO}$ by comparison to
CO intensity maps.

We describe the multi-wavelength datasets used in this work in \S2,
and our method for estimating $X_{\rm CO}$ in \S3. We test the robustness
of our method to systematic uncertainties, by changing the
assumed value for the slope of the molecular SFL, using different SFR
indicators, and CO maps from different telescopes (both
single-dish and interferometers). Our results
are presented in \S4, where we present the observed $X_{\rm CO}$ radial
profile of NGC 628 and compare it to that derived
from dust emission modeling using {\it
  Spitzer}+{\it Herschel}
photometry (Karin Sandstrom private communication). In \S5 and \S6 we discuss the possible physical
origins for the observed spatial distribution in $X_{\rm CO}$, including
the roles of metallicity, gas surface density, and the local UV
radiation field. Finally we provide our conclusions in \S7. For NGC 628, we assume an inclination
of $i=8.7^{\circ}$ and a distance of $8.6$ Mpc \citep{herrmann08}.

\section{Data}

\subsection{VENGA Integral Field Spectroscopy}

All the measurements of nebular emission lines used to estimate
the metallicity, ionization parameter, dust extinction through the Balmer decrement, and
the H$\alpha$ SFR, are made on the
Mitchell Spectrograph (formerly VIRUS-P) IFU data-cube of \mbox{NGC 628} produced by the VENGA survey
\citep{blanc10}. The data-cube samples a rectangular area of
$5.2'\times1.7'$ centered in the nucleus of the galaxy, and has a spatial
resolution of $5.6''$ full-width half-max (FWHM). The spectra covers the 
3555\AA-6790\AA\ wavelength range with an instrumental spectral resolution of
$\simeq5$\AA\ FWHM. At the assumed distance of NGC 628
the VENGA spatial resolution corresponds to
$\sim$235 pc, a few times larger than the typical sizes of large giant
molecular clouds \citep[$\lesssim 60$ pc][]{heyer01} and individual HII
regions ionized by single clusters \citep[$\lesssim 120$ pc,][]{gutierrez11}.
The observations, data reduction, and calibrations will be described
in an upcoming publication.

\subsection{GALEX far-UV and Spitzer MIPS 24 $\mu$m Data}

To ensure that our results are robust against the choice of the SFR
tracer used, along with the VENGA H$\alpha$ derived SFRs, we use a
linear combination of far-ultraviolet (FUV) and mid-infrared (mid-IR)
24$\mu$m emission. The former traces
unobscured star formation, while the latter recovers dust obscured star
formation by means of reprocessed UV radiation reemitted as thermal IR
emission from heated interstellar dust grains. A number of calibrations have been proposed
to use linear combinations of obscured and unobscured tracers to
estimate the SFR \citep{calzetti07, leroy08, kennicutt09, hao11,
  murphy11, leroy12}.

We use the GALEX FUV image of NGC 628 from the GALEX Nearby Galaxy
Survey \citep{gildepaz07} which samples the 1350-1750 \AA\ wavelength
range, has a point-spread-function (PSF) FWHM of 4.5'', and is deep enough for us to measure
the FUV flux at high signal-to-noise over the whole area of interest.
To trace the IR dust emission we use the MIPS 24$\mu$m data of NGC 628,
taken as part of the Spitzer Infrared Nearby Galaxy Survey
\citep[SINGS,][]{kennicutt03} and the Local Volume Legacy survey
\citep[LVL,][]{dale09}. The map has a PSF FWHM of $6''$ and like
the FUV map, it is deep enough for us to measure the 24$\mu$m flux
over the whole area of interest in this paper. The data processing of
both the FUV and 24$\mu$m maps, including the suppression of
backgrounds and the masking of foreground stars is detailed in \cite{leroy12}.

\subsection{HERACLES, BIMA-SONG and CARMA CO Data}

We use three independent datasets to measure $I(CO)$ across the disk
of NGC 628. The Heterodyne Receiver Array CO Line Extragalactic Survey
\citep[\mbox{HERACLES,}][]{leroy09} CO(2-1) map, obtained at the IRAM 30-m single-dish
telescope, has a beam size of $13.6''$ and rms noise of $\sim22$ mK
per 2.6 km s$^{-1}$ channel, which
translates into a  1$\sigma$ limit on the molecular
gas surface density of $\Sigma_{H_2}\sim 3$M$_{\odot}$
pc$^{-2}$. Following \cite{schruba11} we
assume a constant CO(2-1)/CO(1-0) line ratio of 0.7 to estimate I(CO)
from the HERACLES data. 

The Berkeley Illinois Maryland Array (BIMA)
Survey of Nearby Galaxies \citep[\mbox{BIMA-SONG,}][]{helfer03}
CO(1-0) map combines zero spacing single dish data from the
NRAO 12 m telescope and interferometric BIMA C and D array data, resulting in a map with a robust beam size of
$6.2''$, and rms noise of 51 mJy beam$^{-1}$
in a 10 km s$^{-1}$ channel or $\Sigma_{H_2} \sim13$ M$_{\odot}$
pc$^{-2}$. 

\begin{figure}
\begin{center}
\epsscale{1.2}
\plotone{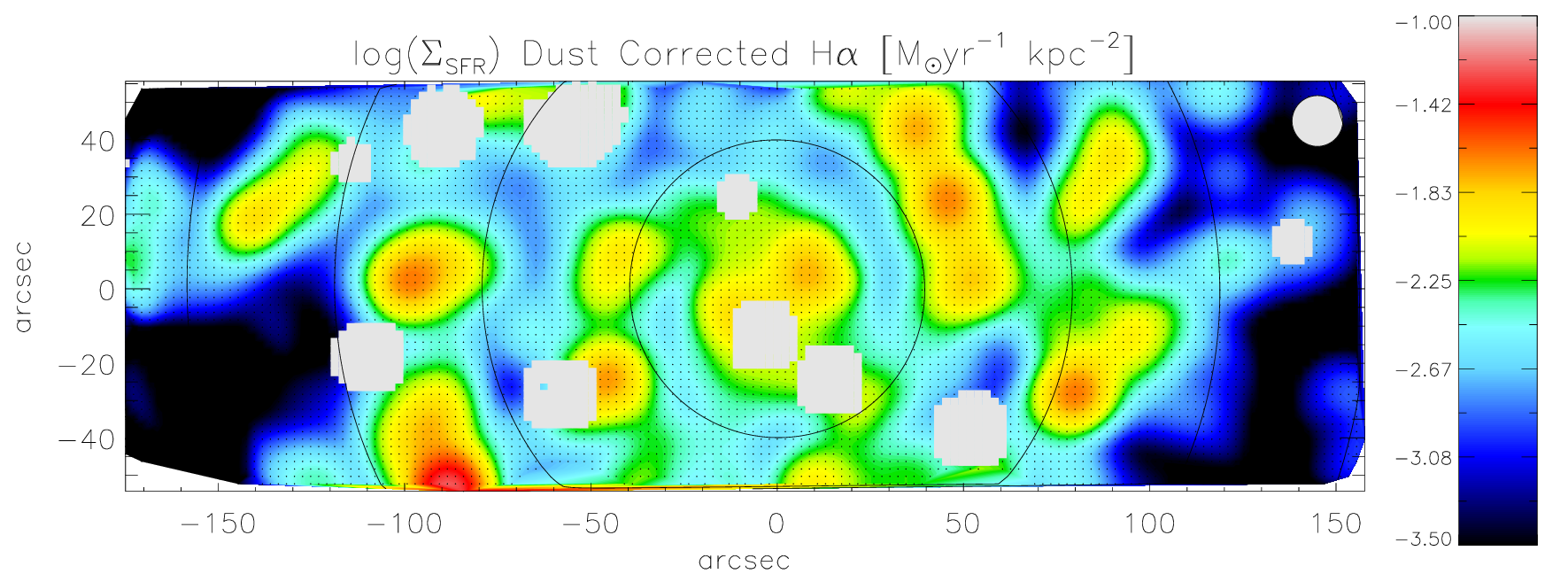}
\plotone{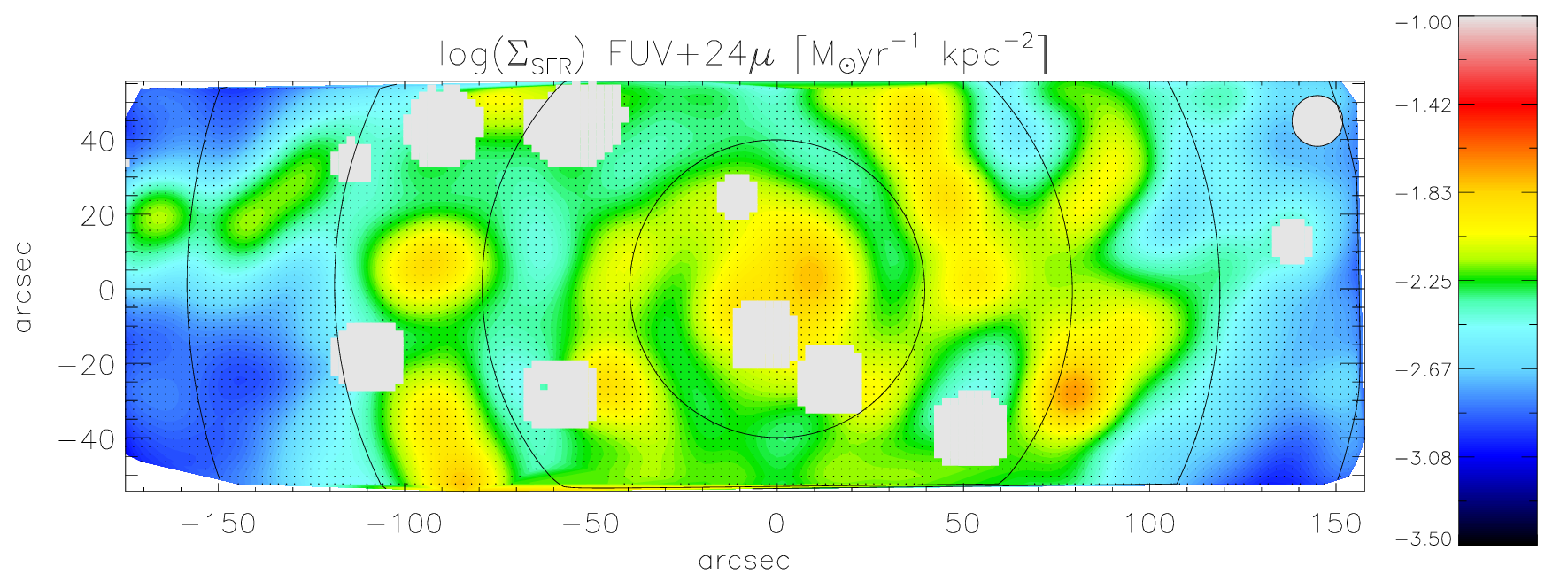}
\plotone{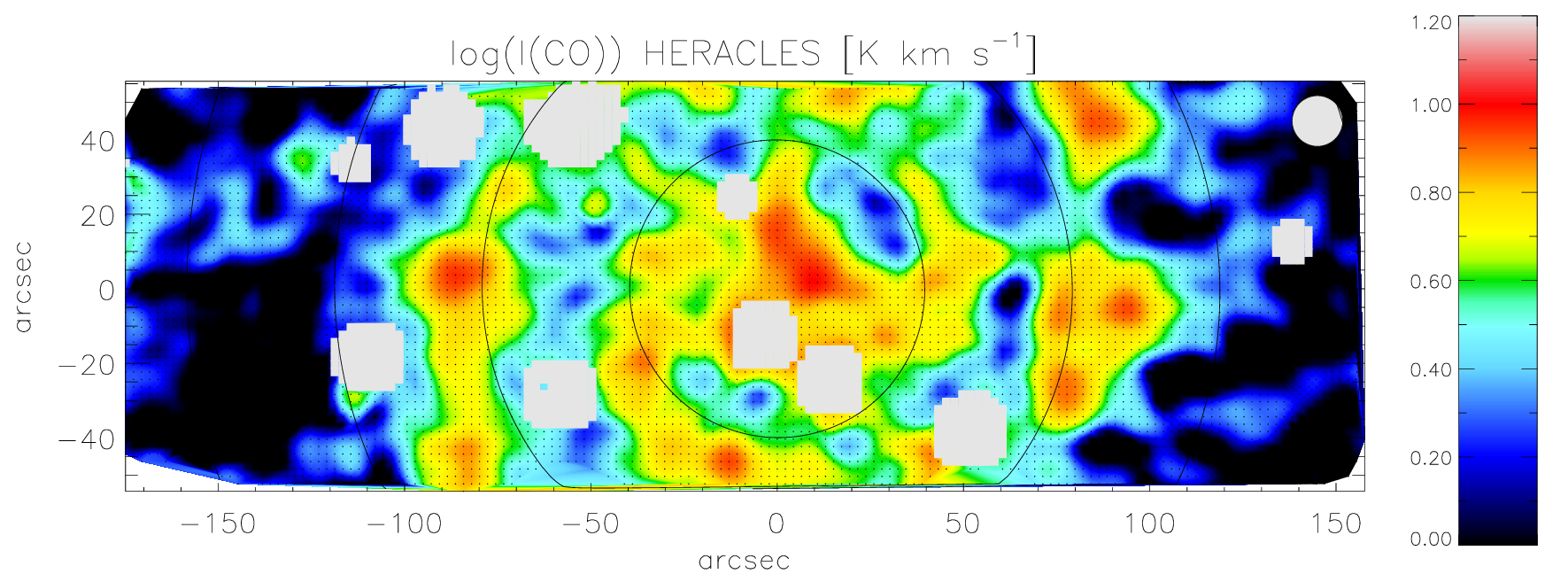}
\plotone{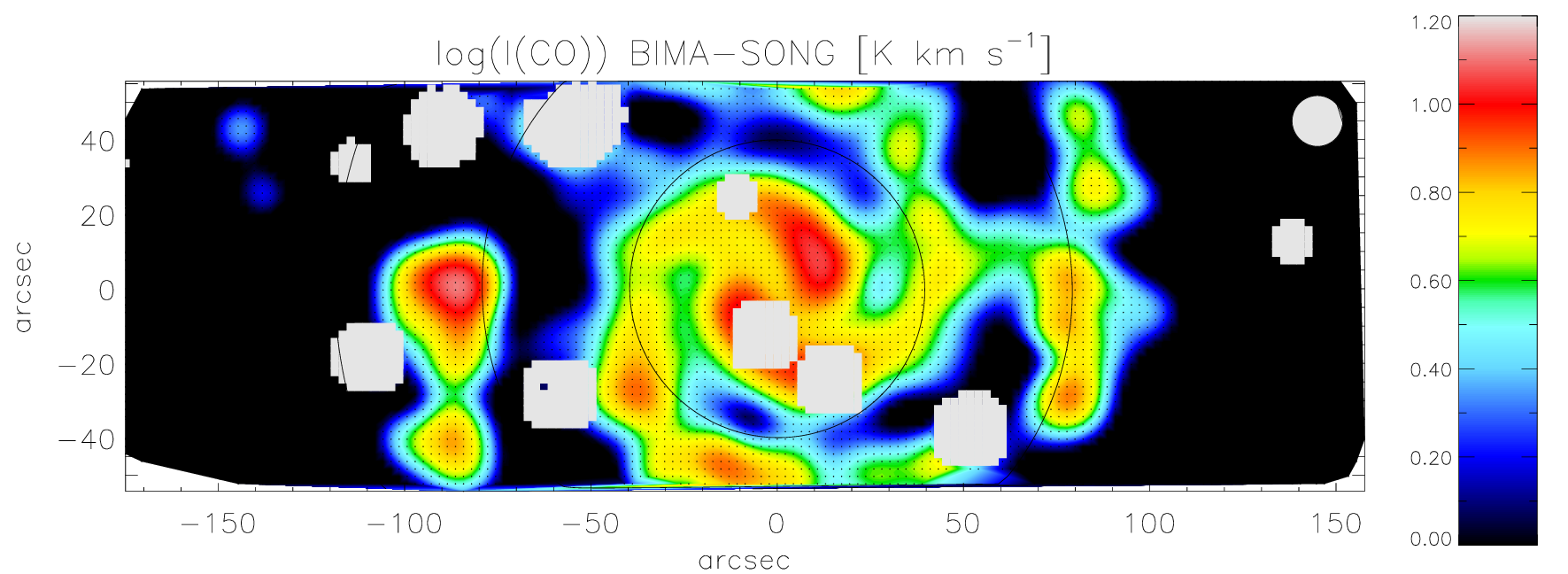}
\plotone{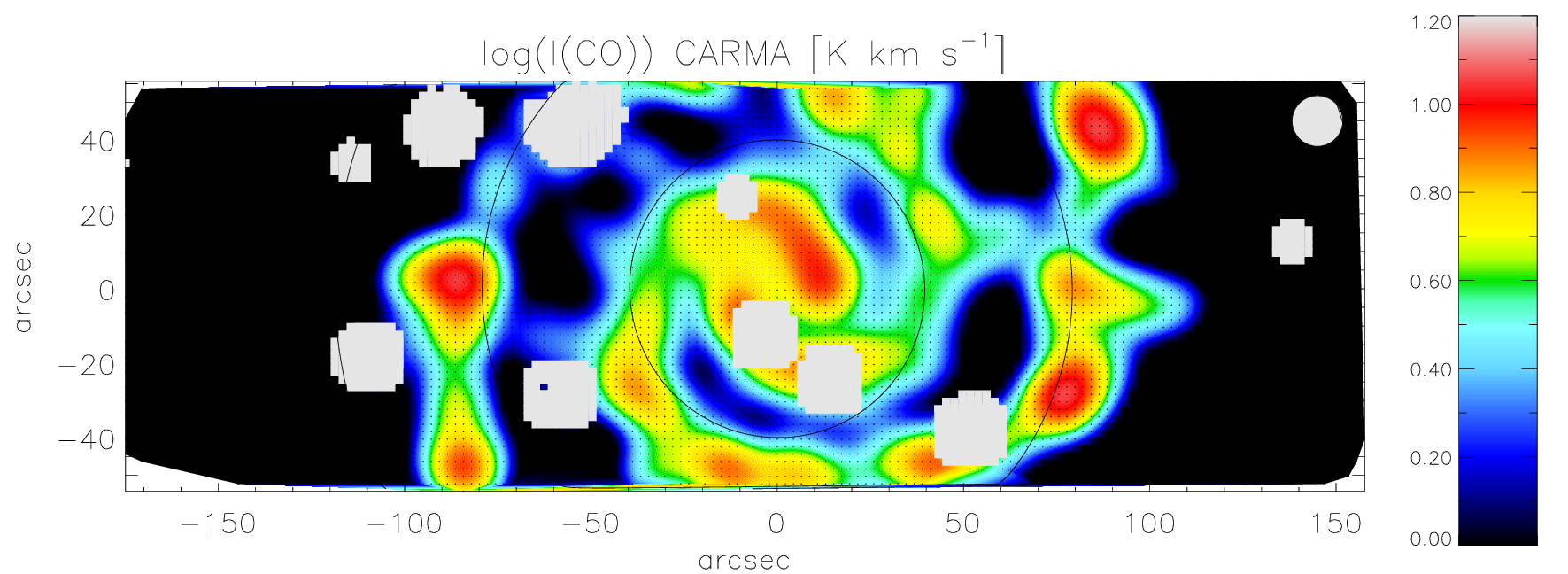}
\caption{Maps of the H$\alpha$, from VENGA, and FUV+24$\mu$m, from
  {\it Galex} and {\it Spitzer-IRAC}, SFR surface density
  (top two), and the CO intensity from the HERACLES, BIMA-SONG, and
  CARMA data respectively (bottom three). All maps are convolved to
  the HERACLES beam-size of $13.6''$ or 570 pc at the distance of NGC
  628 (shown as a white circle in the
  top right corner of each map). Solid lines show the edges of
  the radial annuli in which we measure $X_{\rm CO}$. Gaps in the maps
  correspond to regions contaminated by foreground MW stars.}
\label{fig-1}
\end{center}
\end{figure}

Finally, the CARMA CO(1-0) map of NGC 628 \citep{rahman12}
has a spatial resolution of $3.6''$ and rms noise of $\sim$20 mJy beam$^{-1}$
in a 10 km s$^{-1}$ channel
($\Sigma_{H_2}\sim 5$M$_{\odot}$ pc$^{-2}$). We convolve the BIMA-SONG and CARMA
maps with a Gaussian kernel to match the $13.6''$ PSF of the
HERACLES map, which corresponds to $\sim570$ pc at the assumed
distance to \mbox{NGC 0628}. The three PSF matched CO maps are shown in Figure
\ref{fig-1}.

\begin{figure}[b]
\begin{center}
\epsscale{1.22}
\plotone{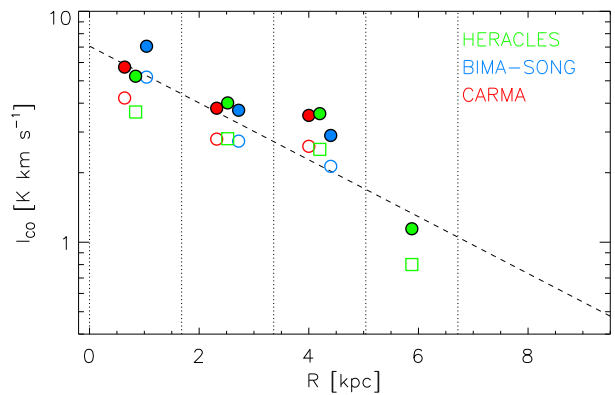}
\caption{CO intensity as a function of radius for the four radial bins
adopted in this work. Filled circles correspond to the CO(1-0)
intensities after applying the scalings described in \S2.3. Open
circles show the raw BIMA-SONG and CARMA measurements before scaling
them to match the HERACLES absolute flux level. Open squares show the
raw CO(2-1) intensity in the HERACLES map before correcting for the
assumed CO(2-1) to CO(1-0) ratio of 0.7. Datapoints within each radial
bin have been shifted in the horizontal direction for
clarity. Vertical dotted lines mark the edges of each radial
bin. The error-bars for each measurement, calculated as the r.m.s. per
beam in each map divided by the square root of the ratio between the
area of each radial bin to the beam-size, are smaller than the
datapoints themselves, and not shown. Differences in the profiles are
dominated by systematic uncertainties at the 20\% level. }

\label{fig-2}
\end{center}
\end{figure}
 
In order to account for systematic errors in the absolute flux
calibration of the data-cubes, and in the assumed CO(2-1) to
CO(1-0) conversion factor, we scale the maps to match the total
CO(1-0) luminosity of the HERACLES map in the inner 5 kpc of the galaxy. We find scaling
factors of 1.36 and 1.36 for the BIMA-SONG and CARMA maps
respectively. Figure \ref{fig-2} presents both the scaled and original
flux measurements in each of the radial bins over which we will
measure $X_{CO}$ (see \S3.2) for the three maps. The
$\sim$30\% difference implied by these scaling factors is consistent with the
systematic uncertainties in the absolute flux calibration of the three
datasets, and the observed scatter in the CO(2-1) to CO(1-0) ratio across the HERACLES
galaxies \citep{leroy09}. Assuming that the CO(2-1) to CO(1-0) ratio
has no significant radial dependance, these corrections only introduce a scaling
in the derived $X_{\rm CO}$ factors. Since we are interested in measuring
relative changes in $X_{\rm CO}$ within the galaxy, a scaling of this type
does not affect our results.

The radial dependance in the CO(2-1) to CO(1-0) ratio across
the disk of spiral galaxies has been studied by \cite{leroy09} by
comparing the CO(2-1) emission in the HERACLES maps to the Nobeyama
45m single-dish CO(1-0) maps of \cite{kuno07} (see Figure 34 in
\cite{leroy09}). Although the scatter seen in the line ratio is large, no
evidence for a radial dependance is seen outside the very central
parts of the galaxies ($R<0.05R_{25}$). Furthermore, as will be
discussed in \S4, the relatively good consistency between the
different datasets (see also Figure \ref{fig-2}), and the agreement
between our measured $X_{CO}$ radial profiles and independent
measurements based on dust modeling indicate that any potential radial
trend in the CO(2-1) to CO(1-0) ratio is at a significantly smaller
level than the radial trends seen in $X_{CO}$.

The consistency between the different measurements shown in Figure
\ref{fig-2} is reassuring. While the BIMA-SONG map includes both
BIMA interferometric data and NOAO 12m single-dish data, the CARMA map
is constructed without the addition of zero spacing information, and
therefore lacks sensitivity on large scales. The agreement seen in
Figure \ref{fig-2} implies that even in the outermost radial bin for
which we can measure the flux in the CARMA map, the lost 
large scale extended emission does not contribute significantly to the
total CO surface brightness.\\

\section{Estimating $X_{\rm CO}$}

Estimating $X_{\rm CO}$ (Equation 1) requires an independent measurement of the
molecular gas column density, or equivalently its surface density ($\Sigma_{H2}$), and the \mbox{CO(1-0)}
line intensity ($I(CO)$). The latter is directly measured from the CO
maps described in \S2.3. We estimate $\Sigma_{H2}$ from the SFR
surface density ($\Sigma_{SFR}$) by inverting the molecular gas star
formation law \citep{kennicutt98, kennicutt07, bigiel08}, which we
parametrize as

\begin{equation}
\frac{\Sigma_{SFR}}{1{\rm M_{\odot}yr^{-1}kpc^{-2}}} =
A\left(\frac{\Sigma_{H2}}{10{\rm M_{\odot}pc^{-2}}}\right)^N
\end{equation}

This is a more general approach than assuming a
constant molecular gas depletion timescale \citep{schruba12, genzel12,
mcquinn12}, which is equivalent to the above method for the particular
case of $N=1$. 

For simplicity, we fix the SFL normalization to
\mbox{$A=-2.3$}, which is equivalent to a molecular gas depletion timescale of 2 Gyr at $\Sigma_{H2}=10$
M$_{\odot}$ pc$^{-2}$. This value is in good agreement with
observations of solar metallicity spiral galaxies in the local universe
\citep[e.g.][]{bigiel08, leroy08, blanc09, schruba11}, and modifying it
introduces a simple scaling in the derived $X_{\rm CO}$ values. We report
the $X_{\rm CO}$ radial profile for two assumed values for the SFL slope of $N=1.0$ and
$N=1.5$, which span the range of plausible slopes allowed
by random and systematic uncertainties in current measurements of the
molecular gas SFL \citep[][ and references within]{calzetti12, kennicutt12}

It is important to explicitly state the assumptions underlying our
method to measure $X_{\rm CO}$. Mainly, we are assuming the existence of a
fundamental power-law like correlation between $\Sigma_{H2}$ and
$\Sigma_{SFR}$, which holds for averaged measurements of these two
quantities over kpc scales. The existence of such correlation across
the disks of spiral galaxies has ben well stablished by previous studies
\citep{kennicutt07, bigiel08, blanc09, verley10, onodera10, liu11,
  rahman12}. 

A potential caveat arises from the fact that the
observed SFL in all the above mentioned studies has been measured by
assuming a constant $X_{\rm CO}$ factor to transform CO surface brightness
into $\Sigma_{H2}$. We do not consider this to be an important
limitation, as reasonable changes in $X_{\rm CO}$ (i.e. of the magnitude
expected from theoretical models for the range of ISM physical properties
present in galaxies) are not large enough as to break the observed
correlation between $\Sigma_{H2}$ and $\Sigma_{SFR}$. On the contrary,
a recent study by \cite{narayanan12} shows that using a changing
$X_{\rm CO}$, which depends on the metallicity and surface density fo the
molecular gas, translates into a smaller scatter around the best-fit
power-law SFL for integrated measurements of galaxies, than when a
single or a bimodal $X_{\rm CO}$ is used. Therefore, a power-law
like SFL on kpc scales appears to exist independently of the
assumption of a constant or changing $X_{\rm CO}$ factor.

A second assumption made when applying our method is that the adopted SFL does
not change across the disk of the galaxy. In the linear SFL scenario
(i.e. when assuming $N=1$) this is equivalent to assuming a constant
depletion timescale across the disk of the galaxy. In the linear case,
our method cannot break the degeneracy between the measured values for
$X_{\rm CO}$ and the assumed gas depletion timescale. As mentioned above,
the absolute adopted value only introduces a simple scaling of
$X_{\rm CO}$, which is not important given that we are interested in
studying relative changes across the disk of the galaxy. However, any
observed changes in $X_{\rm CO}$ within the galaxy could, in principle, be
attributed instead to changes in the depletion timescale, or
equivalently, in the star formation efficiency (SFE, defined here as
the inverse of the depletion time). In \S4.3 we discuss
this possibility, only to conclude that it is very unlikely that
changes in the depletion time can explain the observed $X_{\rm CO}$ radial
profile, and that the observed change in the conversion factor is
most likely real. In the non-linear case ($N=1.5$), the
depletion timescale, or equivalently the SFE, does
change as a function of the local surface density of molecular gas.

\subsection{Measurement of the SFR}

We produce maps of $\Sigma_{SFR}$ using two independent SFR indicators. First
we use the H$\alpha$ emission line flux, corrected for dust extinction
using the Balmer decrement. The methods used to measure dust
corrected H$\alpha$ fluxes and to correct the emission line map for
the contribution from diffuse ionized gas (DIG) are analog to the ones
described in \cite{blanc09}. An updated description of the methods
used, including slight differences in the methodology used for flux
calibration and construction of the IFU data-cubes will be described
in an upcoming publication. We use the H$\alpha$ SFR calibration of
\cite{murphy11} and \cite{hao11}, taken from the compilation in
\cite{kennicutt12}, which assumes a \cite{kroupa03}
initial-mass-function (IMF). Note that this differs from the \cite{salpeter55}
IMF used in \cite{blanc09}. 

The second SFR indicator used is a linear combination of FUV and
24$\mu$m flux. We use the calibration of \cite{leroy12}, which assumes
a \cite{chabrier03} IMF. Calibrations based on the Kroupa and Chabrier IMFs yield nearly
identical SFRs \citep{chomiuk11}, which are typically $\sim$30\% lower
than those obtained assuming a Salpeter IMF. Since H$\alpha$ and FUV
photons trace star formation over different timescales (roughly 10 and 100 Myr
respectively), systematic differences between the two independent
estimates might arise as a consequence of differences in star
formation history. Analogously to what was done with the CO maps, we
attempt to remove any systematic differences in the flux calibration
of the different datasets and the SFR calibrations used, by matching
the total SFR within the central 5 kpc of the galaxy. This translates
in us multiplying the $\Sigma_{SFR, FUV+24\mu m}$ map by a factor of
0.8 to match the $\Sigma_{SFR, H\alpha}$ map. Again, this correction does
not affect the relative differences in $X_{\rm CO}$ which we are trying to
measure, but only introduces a scaling of the derived values across
the whole galaxy. Both maps are convolved with a Gaussian kernel in order
to match the common $13.6''$ FWHM PSF, and are presented in Figure
\ref{fig-1}.

\subsection{Calculation of $X_{\rm CO}$}

We use the PSF matched SFR maps to measure $\Sigma_{SFR, FUV+24\mu m}$,
$\Sigma_{SFR, H\alpha}$, and $I(CO)$ in elliptical annuli of
constant galactocentric radius. We chose an annuli width of $40''$
which roughly corresponds to three times the spatial FWHM of the maps,
and a physical scale of 1.68 kpc at the adopted distance to NGC
628 (see Figure \ref{fig-1}). The sensitivity of the BIMA-SONG and
CARMA maps only allows the measurement of $I(CO)$ out to 5 kpc ($\sim
120''$) which corresponds to the inner three annuli, while the
HERACLES map is able to provide a reliable measurement in a fourth
bin, out to a galactocentric distance of 7 kpc ($\sim170''$). While
\cite{schruba11} was able to measure CO emission using the HERACLES
map of NGC 628 out to larger radii ($\sim$10 kpc) using a stacking
technique, here we are also limited by the coverage of the VENGA IFU
observations which only reach out to $\sim$7.5 kpc. We use the
extended stacking measurements of \cite{schruba11} separately in the
next section to extend our measurements to larger radii. Having
measured these three quantities ($\Sigma_{SFR, FUV+24\mu m}$,
$\Sigma_{SFR, H\alpha}$, and $I(CO)$), we use Equation 2 to obtain
$\Sigma_{H2}$, and we transform it to units of column density. Finally we 
calculate $X_{\rm CO}$ for each radial bin by applying Equation 1. The central radius and measured $X_{\rm CO}$ values
for each annuli are reported in Table \ref{tbl-1}.

The uncertainty in the measured $X_{\rm CO}$ values is dominated by the
intrinsic scatter in the molecular SFL. A series of studies have shown
that this quantity depends on the physical scales over which the
observed surface densities are integrated \citep{verley10, onodera10,
schruba11,  liu11}. For the large physical scales considered in this work
($\sim4$ kpc given the mean area of our radial bins), we assume a value
of $\sim0.15$ dex for the scatter, which we propagate into the
uncertainty in $X_{\rm CO}$. Thanks to the high $S/N$ of the datasets and
the large areas over which we are integrating, the contribution from
random errors in $\Sigma_{SFR}$, which includes photometric errors in
the $H\alpha$ and $H\beta$ (the latter entering in the Balmer
decrement dust extinction correction), and in the FUV and 24 $\mu$m
fluxes, as well as from noise in the CO maps, is negligible (a factor
of $\sim5$ smaller) compared to the scatter in the molecular SFL.

\section{The $X_{\rm CO}$ Radial Profile}

\subsection{Consistency Between Different Datasets}

\begin{figure*}
\begin{center}
\epsscale{1.1}
\plotone{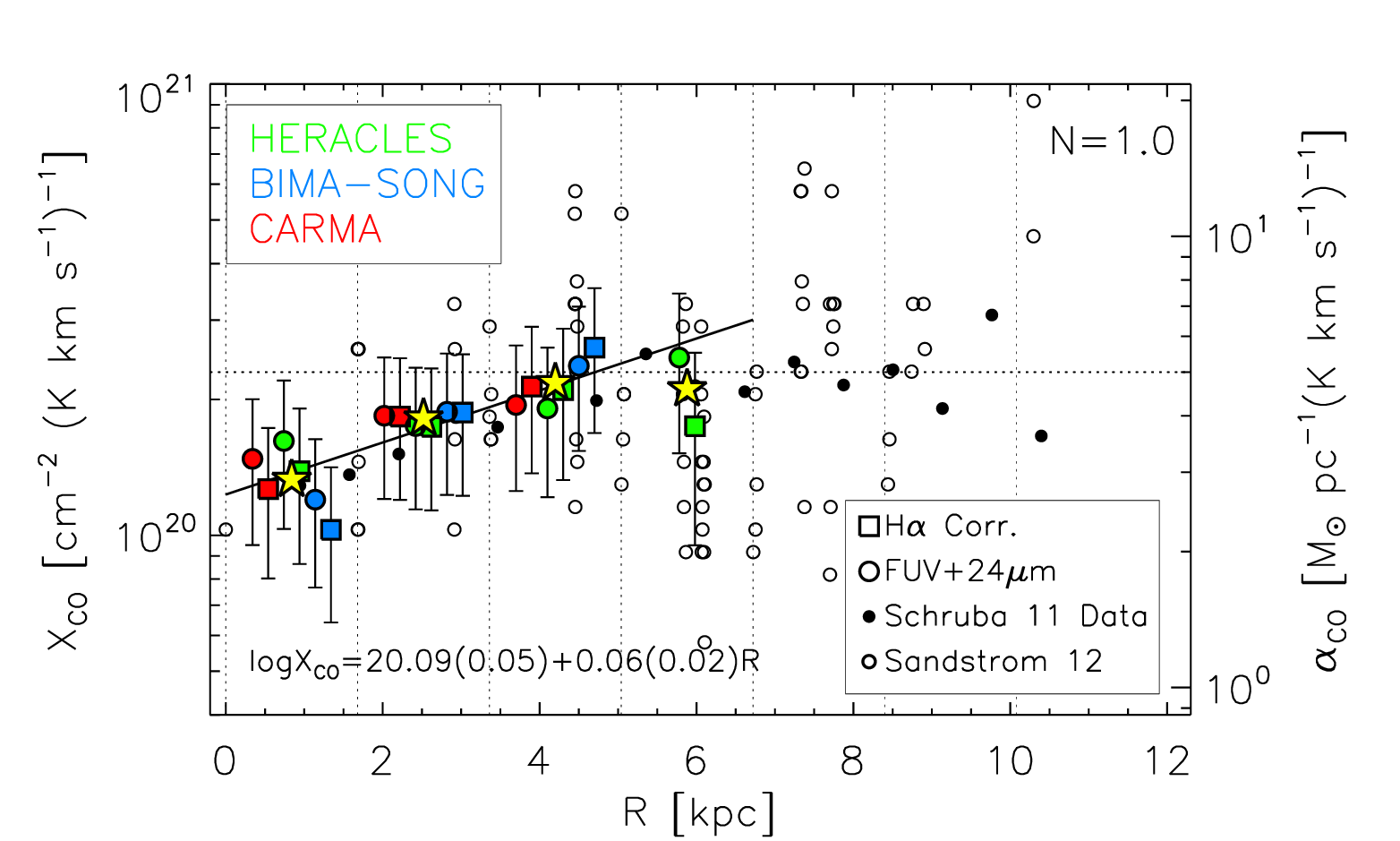}
\plotone{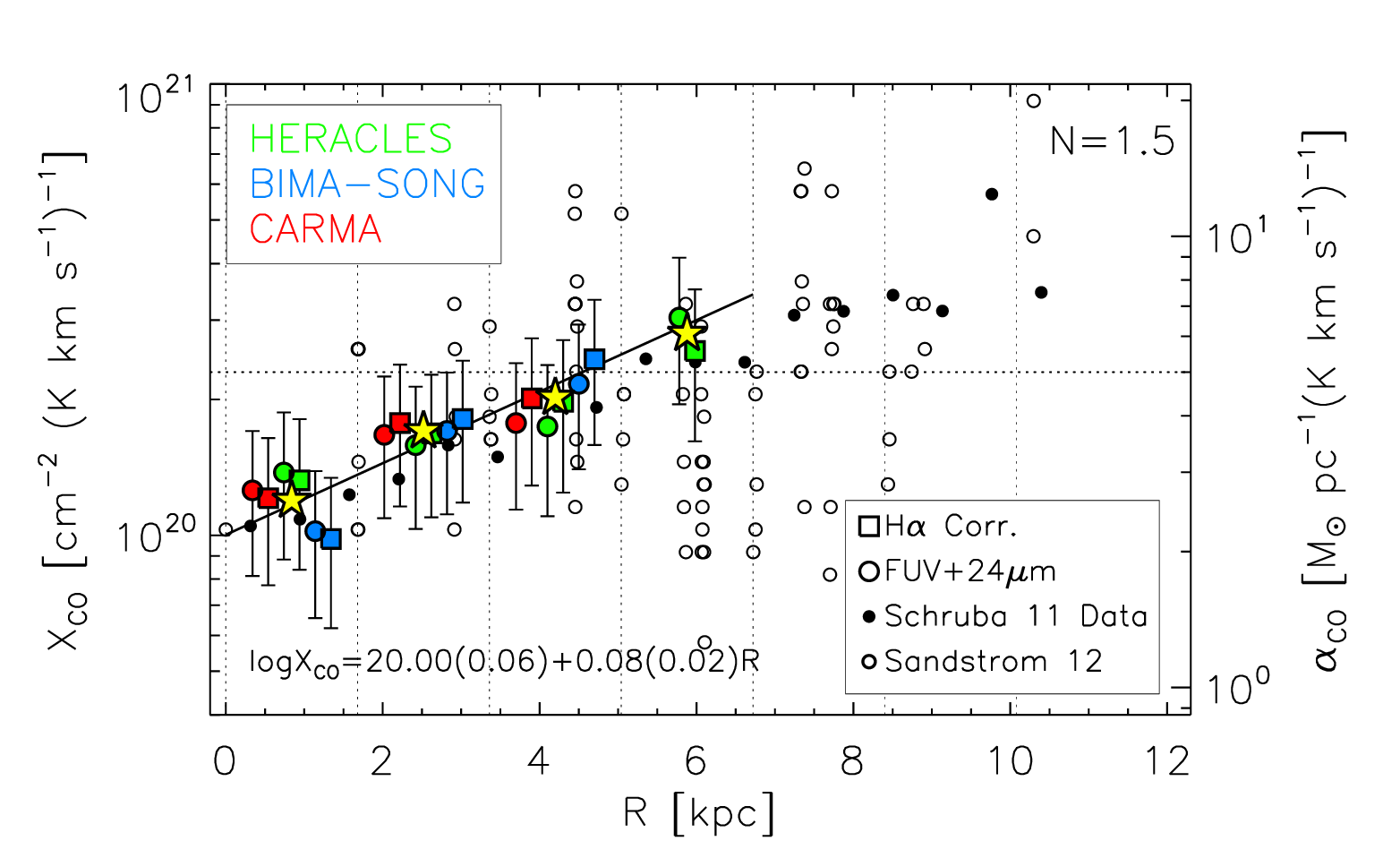}
\caption{Radial profile of $X_{\rm CO}$ in NGC 628. The top and bottom panels present
results for a SFL slope of $N=1.0$ and $N=1.5$
respectively. Measurements using the HERACLES, BIMA-SONG, and CARMA
maps are shown in green, blue, and red respectively. Measurements
based on H$\alpha$ and FUV+24$\mu$m SFRs are shown as large color
squares and circles respectively. Datapoints within each radial bin have been shifted in the
horizontal direction for clarity. Yellow stars show the average of
these measurements in each radial bin. Vertical dotted lines mark the
edges of each radial bin, and the horizontal dotted line marks the
MW $X_{\rm CO}$ factor of \cite{pineda10}. Also shown are measurements
from {\it Spitzer}+{\it Herschel} dust SED modelling (Karin Sandstrom
private communication, small open circles), and the results of
applying our method to the stacked data in \cite{schruba11} (small filled circles).} 
\label{fig-3}
\end{center}
\end{figure*}

In Figure \ref{fig-3} we present the measured $X_{\rm CO}$ factor as a function
of galactocentric radius for the six combinations of datasets (two SFR
indicators and three CO maps from different telescopes), and two
different assumptions for the molecular SFL slope ($N=1.0$ and
1.5). At all radii, all combinations of datasets are consistent with
each other within the 1$\sigma$ formal error bars, and show the same
systematic trend of increasing $X_{CO}$ with radius. This implies that our
results are robust against potential systematics associated with the adopted
SFR indicator, and the nature of the CO maps utilized (single-dish
vs. interferometer, CO(1-0) vs. CO(2-1)). The consistency between
different datasets also confirms that photometric errors are not a significant source of
uncertainty, when compared to the systematic errors associated with
the scatter in the SFL. 

As can be seen in Figure \ref{fig-1}, the agreement between the
H$\alpha$ and FUV+24$\mu$m SFR maps is good. After performing the
scaling described in \S3.1, the ratio of the two different SFRs
measured in the inner four radial bins shows a r.m.s. scatter of 13\%.
When analyzing the morphology of these two SFR tracers we
notice a strong similarity. All individual star forming regions picked
up by one method are also recovered by the other. This is important as
it implies that the dust corrected H$\alpha$ method is not missing a
significant fraction of completely obscured star formation which
should be recovered by dust emission at 24$\mu$m. 

While both maps trace the same structures, a more detailed inspection shows that the
FUV+24$\mu$m SFR map does seems to have a more diffuse morphology than
the H$\alpha$ SFR map. In particular more diffuse FUV+24$\mu$m emission is seen
in the inter-arm regions of the galaxy, while the H$\alpha$ emission
seems to be more concentrated in the arms. This is consistent with the
typical star formation timescales traced by these two
methods. H$\alpha$ traces young stars ($\leq 10$ Myr), which are seen
close to their birthplaces, while both FUV and 24$\mu$m emission
receive a significant contribution from older stars ($\leq 100$ Myr)
which are more homogeneously distributed across the galactic disk, and
not necessarily associated with the spiral arms.

 The morphological agreement between the different CO maps is also very
good. After applying the scalings described in \S2.3 the CO
luminosities in the inner 3 radial bins show a typical r.m.s. scatter
of 20\% (see Figure \ref{fig-2}). The agreement justifies our adoption of a single CO
(2-1)/(1-0) line ratio across the galaxy. 
It is interesting to notice that while that both the CO emission
and the SFR trace similar morphological structures across the galaxy,
that is mainly the grand design spiral arms of NGC 628, a
detailed comparison shows that there is not a one to one correlation
between individual CO bright regions and regions which are bright in
H$\alpha$ or FUV+24$\mu$m. This reflects the temporal dependance of
the star formation process, in which recently formed star clusters
disrupt and dissociate their molecular gas birth clouds by injecting
both radiative and mechanical energy into their surrounding ISM. As
this process takes place, and individual star forming region evolves
from a CO bright molecular complex into a 24$\mu$m bright, highly
obscured, star forming region, which eventually dissociates and removes
the bulk of the molecular gas and dust surrounding it, becoming a
H$\alpha$ and FUV bright region in the process. This temporal
evolution translates into an spatial offset between SFR and molecular
gas tracers, and is the main cause behind the need to average over
large kpc scales in order to do a measurement like the one presented
in this work \citep{schruba10, madore10, feldmann12b, kennicutt12}

In order quantify the previous statements regarding the robustness of
our measurement we calculate the standard deviation of
all $X_{\rm CO}$ measurements in each radial bin, and find it to range from a
few percent to a $\sim25$\% of the measured values in the worst case
(outermost bin). The mean r.m.s. across all four bins
corresponds to $\sim15$\% of the measured values. This dispersion
is caused by a combination of different factors including systematic
errors in the adopted CO(2-1)/CO(1-0) ratio, a possible systematic
lack of sensitivity to low surface brightness emission in the
interferometric CO maps, and potential variations in the
ratio between the H$\alpha$ and FUV+24$\mu$m SFRs which might arise as
the consequence of changes in the stellar populations and the
properties of dust across the disk of the galaxy. The magnitude of
these systematic deviations is smaller than the formal 0.15 dex ($\sim35$\%)
uncertainty adopted. Nevertheless, we add this dispersion in
quadrature to our formal errors in order to take into account the
effects of theses systematics in our error budget.

\begin{deluxetable*}{ccccccccc}

\tabletypesize{\scriptsize}

\tablecaption{NGC 628 $X_{CO}$ Radial Profile\label{tbl-1}}

\tablewidth{0pt}

\tablehead{
\colhead{$R_{bin}$} & 
\colhead{$N$} &
\colhead{} &
\colhead{$X_{CO}$ - H$\alpha$\tablenotemark{a}} &
\colhead{} &
\colhead{} &
\colhead{$X_{CO}$ - FUV+24$\mu$m\tablenotemark{a}} &
\colhead{} &
\colhead{$\langle X_{CO} \rangle$\tablenotemark{a}}
}

\startdata
&\ \ \  & HERACLES & BIMA & CARMA & HERACLES & BIMA & CARMA & \\ 
\tableline\\
0.84  &   1.0 & 1.4$\pm$0.5  & 1.0$\pm$0.4 & 1.3$\pm$0.5 & 1.6$\pm$0.6 & 1.2$\pm$0.4 & 1.5$\pm$0.5 & 1.3$\pm$0.5 \\
2.52  &         & 1.7$\pm$0.6  & 1.9$\pm$0.7 & 1.8$\pm$0.6 & 1.7$\pm$0.6 & 1.9$\pm$0.7 & 1.8$\pm$0.6 & 1.8$\pm$0.6 \\
4.20  &         & 2.1$\pm$0.8  & 2.6$\pm$0.9 & 2.1$\pm$0.8 & 1.9$\pm$0.7 & 2.4$\pm$0.8 & 1.9$\pm$0.7 & 2.2$\pm$0.8 \\
5.88  &         & 1.7$\pm$0.8  & -                    & -                    & 2.5$\pm$1.0 & -                   & -                    & 2.1$\pm$0.9 \\
\tableline\\
0.84  &   1.5 & 1.3$\pm$0.5  & 1.0$\pm$0.4 & 1.2$\pm$0.4 & 1.4$\pm$0.5 & 1.0$\pm$0.4 & 1.3$\pm$0.5 & 1.2$\pm$0.4 \\
2.52  &         & 1.7$\pm$0.6  & 1.8$\pm$0.6 & 1.8$\pm$0.6 & 1.6$\pm$0.6 & 1.7$\pm$0.6 & 1.7$\pm$0.6 & 1.7$\pm$0.6 \\
4.20  &         & 2.0$\pm$0.7  & 2.5$\pm$0.9 & 2.0$\pm$0.7 & 1.7$\pm$0.6 & 2.2$\pm$0.8 & 1.8$\pm$0.6 & 2.0$\pm$0.8 \\
5.88  &         & 2.6$\pm$1.0  & -                    & -                    & 3.0$\pm$1.1 & -                   & -                    & 2.8$\pm$1.0 \\

\enddata

\tablenotetext{a}{In units of 10$^{20}$ cm$^{-1}$(K km s$^{-1}$)$^{-1}$}

\end{deluxetable*}

\subsection{Dependance of $X_{\rm CO}$ on the Assumed SFL Slope.}

The dotted horizontal line in Figure \ref{fig-3} marks the canonical
Milky Way $X_{\rm CO}$ factor of $2.3\times10^{20}$ cm$^{-1}$(K km s$^{-1}$)$^{-1}$ from
\cite{pineda10}, which is adopted as the preferred value in \cite{kennicutt12}.
While at radii $>2$ kpc the $X_{\rm CO}$ factor in NGC 628 is consistent
with the MW value, there is an evident drop in the central
2 kpc of the galaxy, where the measured $X_{\rm CO}$ is a factor of 2 lower
than the MW value. 
 
The drop in the radial profile of $X_{\rm CO}$ towards the central
regions is observed independently of the assumed SFL slope. 
A non-linear SFL with $N>1$ implies lower H$_2$ surface
densities, and therefore lower derived $X_{\rm CO}$ values for high SFR
surface density regions than a linear relation. Since the SFR in \mbox{NGC
628}, and in most disk galaxies, decreases with radius, a non-linear
slope implies an even larger drop in $X_{\rm CO}$ towards the central
regions. In any case, the magnitude of the changes in $X_{\rm CO}$ between the $N=1.0$
and 1.5 cases is smaller than the changes in $X_{\rm CO}$
observed as a function of radius across the galaxy. This implies that
potential changes in the SFL slope
within the galaxy would not affect our results
significantly. While a sub-linear ($N<1$) slope could reduce the
magnitude of the observed drop in $X_{\rm CO}$ towards the central
regions, this scenario is not currently supported by observations
\citep[][and references within]{calzetti12, kennicutt12}.

This dependance with the SFL slope is
important, as it implies that the observed radial trend of $X_{CO}$ is not a consequence
of changes in the star formation efficiency with
radius. Adopting a super-linear slope for the SFL is effectively
testing the case in which the efficiency of star formation is enhanced
in higher density regions, and as stated above, this effect only
boosts the decrease of $X_{\rm CO}$ at small radii, although not
significantly. We further discuss this subject in the next section.

\subsection{$X_{\rm CO}$ Gradient and Comparison to Other Measurements}

Also shown as open circles in Figure \ref{fig-3} are the measurements of $X_{\rm CO}$ in
NGC 628 from dust SED modelling (Karin Sandstrom private communication). We only plot their data-points with an
uncertainty $\leq 0.4$ dex in $X_{\rm CO}$. This study uses the method of
\cite{leroy11} to simultaneously model the dust mass surface density
and the dust-to-gas ratio across the disks of a sample of nearby
spiral galaxies with far-IR SED measurements from the SINGS and KINGFISH
\citep{kennicutt11} surveys. The dust mass surface density, in
combination with the dust-to-gas ratio, provides an
estimate of the total gas surface density. Subtraction of the atomic
component (measured from HI 21cm maps) yields the
surface density of the molecular component. This technique is completely
independent of the method adopted in this work. The data-points from
the dust SED modelling have a typical uncertainty of 0.2 dex at $R<7$ kpc,
hence they agree with our measurements across the
whole range in radii sampled by our data, and significantly show the
same trend of decreasing $X_{\rm CO}$ towards the central regions of NGC 628. No scaling has been
applied to the measurements in Figure \ref{fig-3},
so the agreement is not only in the shape of the $X_{\rm CO}$ radial
profile but also in its absolute value. 

A similar analysis was conducted by
\cite{aniano12} for NGC 628 and NGC 6946. While a drop in $X_{CO}$
towards the center of the galaxy is evidently seen in their data for
NGC 6946, the authors do not claim the detection of a changing
$X_{CO}$ profile in NGC 628. Inspecting their Figure 4, its is evident
that the results of their measurements of $X_{CO}$ in NGC 628 are
subject to strong systematic uncertainties, and show large changes
(from a decreasing to a flat, or even decreasing radial trend)
depending on which subset of the {\it Spitzer}+{\it Herschel} data is
used and the spatial resolution at which the modeling is
conducted. This is not the case for NGC 6946 where different datasets
at different resolutions yield consistent results. Given the large
systematic uncertainties in their measurement of $X_{CO}$ for NGC 628,
we refrein from conducting a detailed comparison to their results.

In \S3, we discussed the inherit degeneracy in our method between the SFL
normalization constant (or equivalently the gas depletion time when
$N=1$), and the measured values of $X_{\rm CO}$. A basic assumption in
this work is that of a constant SFL across the disk of NGC 628. In
order to explain the data in Figure \ref{fig-3} solely as a variation
in the normalization of the SFL would require the depletion timescale
to be longer, or the star formation efficiency to be lower, by a
factor of 2 in the central (denser) regions than in the outer
disk. This is not in line with expectations from measurements of the
SFE across different environments in galaxies in the local universe. A
series of studies have shown that luminous infrared galaxies (LIRGS),
ultra-luminous infrared galaxies (ULIRGS), and sub-mm galaxies (SMGs),
in which the molecular gas surface density is one to two orders of
magnitude larger than in the disks of normal spirals, show typical
depletion timescales which are factors of 3-4 shorter than normal
spiral galaxies. The difference becomes even larger (factors of 4-10
shorter) if differences in the $X_{\rm CO}$ factor in these systems are
taken into account \citep{daddi10, genzel10,
  garciaburillo12}. Although the central region ($<2$ kpc) of NGC
628 is only factors of a few denser than the outer disk, it would be
surprising to find a reversal from the global trend seen in the SFE
towards denser environments. Furthermore, the consistency between the
results from this work and the independent dust SED modelling method,
which is not subject to such a degeneracy between
$X_{\rm CO}$ and the depletion timescale, is very encouraging, and
gives us confidence that our working assumption is valid within the
range of current uncertainties. 

If we model the $X_{\rm CO}$ profile in the inner 7 kpc of the galaxy as a simple gradient of the form:

\begin{equation}
{\rm log}(X_{\rm CO})={\rm log}(X_{CO,0})+\Delta{\rm log}(X_{\rm CO})\times R
\end{equation}

\noindent
a fit to the average of the six measurements in each radial bin (two
measurements in the outer bin), which are shown as stars in Figure
\ref{fig-3}, yields $X_{\rm CO}$ gradients of $\Delta{\rm
    log}(X_{\rm CO})=0.06\pm0.02$ dex kpc$^{-`}$ for $N=1$, and
\mbox{$\Delta{\rm log}(X_{\rm CO})=0.08\pm0.02$ dex kpc$^{-1}$} for
$N=1.5$. 

The best-fit central $X_{\rm CO}$ values are
$X_{CO,0}=1.3\pm0.2\times10^{20}$ cm$^{-1}$(K km s$^{-1}$)$^{-1}$ for
$N=1$ and $X_{CO,0}=1.1\pm0.2\times10^{20}$ cm$^{-1}$(K km s$^{-1}$)$^{-1}$ for
the $N=1.5$ case. Our data imply that $X_{\rm CO}$ in the galactic center of \mbox{NGC
628}  is about a factor of two higher than the typical values of
$X_{CO,0}=0.1-0.5\times10^{20}$ cm$^{-1}$(K km s$^{-1}$)$^{-1}$
measured in the MW Galactic Center \citep[e.g.][]{sodroski95, strong04, oka98}. 

Inspecting Figure \ref{fig-3}, particularly for the $N=1$ case, it is
not evident if the radial profile of $X_{\rm CO}$ follows a linear
gradient, or if it is fairly flat across the disk of the galaxy, and
only drops in the central regions. Our measurements are limited
at $R>7$ kpc by the sensitivity of the HERACLES CO map, and the
coverage of the VENGA IFU data-cube. The dust SED modelling datapoints
go out to $\sim10$ kpc, but they show a very large
scatter, and at such large radii, the uncertainty in their
measurements increases sharply, so it is not easy to draw conclusions
from these data-points regarding the behavior of $X_{\rm CO}$ at large
radii. 

In an attempt to extend our measurements into the outer disk of the
galaxy, we use our method to estimate $X_{\rm CO}$ on the stacked CO and
FUV+24$\mu$m radial profiles of \mbox{NGC 628} presented in
\cite{schruba11}. In this work, the authors use the 21 cm HI velocity
field to predict the exact wavelength of the CO line at every position
in the HERACLES data-cube. This information is then used to register
and stack the spectra of all spatial resolution elements in $15''$
wide radial bins across the galaxy. Using this technique,
\cite{schruba11} is able to significantly measure $I(CO)$ out to
$\sim10$ kpc. They also measure the FUV+24$\mu$m SFR profile using the
same maps used in this work. We show these measurements as filled
black circles in Figure \ref{fig-3}. Error-bars are not shown for
clarity, but they are also dominated by the intrinsic scatter in the
SFL, and are of the same order of magnitude as the error-bars
for the measurements in the inner regions of the galaxy (color filled circles and
squares). The results of this exercise are inconclusive. While in the
$N=1$ case the value of $X_{\rm CO}$ seems to flatten around the canonical
MW value, if we assume $N=1.5$ $X_{\rm CO}$ such hints for a flattening in
the outer regions are less obvious, but still present.

As will be further discussed in \S5, NGC 628 shows a metallicity gradient described by a fairly
constant slope out to $\sim$1 $R_{25}$ \citep[$\sim$13 kpc][]{rosales-ortega11}. If $X_{\rm CO}$
scales as some power of the metal abundance ($Z$), we would
expect it to keep on increasing out to a similar radius. On the other hand
there are reasons why a flattening in the radial profile might be
expected. For example, if $X_{\rm CO}$ increases with a decreasing
molecular cloud surface density, as proposed by \cite{narayanan12},
and the GMCs in the outer disk have uniform properties but become
denser towards the central regions, a certain level of flattening in
the $X_{\rm CO}$ profile might occur, although the metallicity dependance
discussed above should still be present. We
will explore the role of the molecular gas surface density and the UV
radiation field in \S6, and will discuss this possibility in more
detail. 

For all the results presented in the following sections, we
adopt the $N=1.0$ case. We have checked that changing the slope
to $N=1.5$ does not significantly affect the rest of the results
presented below.\\

\section{Relating the $X_{\rm CO}$ and Metallicity Gradients}

As discussed in \S1, both observations of nearby and high
redshift galaxies \citep{arimoto96, israel97, bolatto08, leroy11, schruba12,
  genzel12} as well as theoretical models \citep{krumholz11, narayanan12, feldmann12}
indicate that $X_{\rm CO}$ is a decreasing function of metallicity. The
main physical process behind this dependance is the role that dust extinction,
which is directly linked with metallicity, plays in the dissociation of
CO molecules. In photo-dissociation regions on the edges of molecular
clouds, as the metallicity and with it the dust extinction
decreases, the CO to C$^+$ transition layer moves inwards, leaving
behind large envelopes of ``CO dark'' gas where hydrogen is still in
molecular form thanks to self-shielding \citep{maloney88, bolatto99, glover10}. This translates into higher CO
to H$_2$ conversion factors for lower metallically environments.
While at the relatively high metallicities found across the disk of NGC
628 we do not expect to see extreme cases of completely ``CO dark''
molecular clouds \citep[which might be present in extreme low
metallicity galaxies like I Zw 18,][]{schruba12}, variations in the CO brightness of the diffuse
lower density (and lower extinction) envelopes of molecular
clouds as function of metallicity are possible \citep{feldmann12}.

Since most disk galaxies show radial metallicity gradients,
with higher metal abundances towards the central regions \citep[][and
references therein]{moustakas10}, it would be
natural to attempt to relate metallicity gradients with 
gradients in $X_{\rm CO}$. In this section we study the relation
between the observed radial profile of $X_{\rm CO}$ in NGC 628, and its
metallicity distribution. We measure the metallicity gradient using
strong nebular emission lines in the VENGA IFU data, and study
how the relation between $X_{\rm CO}$ and metallicity across the disk of
the galaxy compares to that seen across galaxies in the local
universe, and to the predictions of theoretical models.

\subsection{The Metallicity Gradient in NGC 628}

 The VENGA IFU data-cube provides maps of many strong nebular emission
lines typically used for metallicity diagnostics. As discussed in
detail in \cite{kewley08} and \cite{moustakas10},
different strong line methods used to measure metallicity can show large discrepancies in the
derived abundances. Particularly, methods calibrated against
theoretical photo-ionization models typically yield higher
metallicities than methods calibrated against samples of HII regions
with direct electron temperature measurements (a.k.a. empirically
calibrated methods). Differences as large as
0.6 dex are seen between these two families of methods. On the other
hand, relative differences in metallicity from one system to another do
not change drastically if different methods are applied, allowing meaningful
conclusions to be drawn from comparing different samples, as long as a
single method is used, or measurements using different methods are
properly transformed to a common metallicity scale \citep{kewley08}.
In order to study the metallicity dependance of $X_{\rm CO}$  across NGC 628,
and compare it to previous measurements in the literature, we must not
only choose a reliable strong-line abundance indicator, but we must
also put ours and the literature measurements on a common metallicity
scale. 

Our preferred metallicity calibration is the $N2O2$ method of
\cite{kewley02}, which estimates the oxygen abundance from the
[NI]$\lambda$6584/[OII]$\lambda$3727 ratio, and is calibrated against
photo-ionization models. We have a series of
reasons to favor this calibration. Mainly, the $N2O2$ indicator is highly
independent of the ionization parameter, is single valued at all
metallicities, and the secondary production of nitrogen makes this indicator very
sensitive to metallicity changes at high metallicities like the ones
we expect in massive spiral galaxies like \mbox{NGC 628}, \citep[see
discussion in][]{kewley02}.

\begin{figure}
\begin{center}
\epsscale{1.2}
\plotone{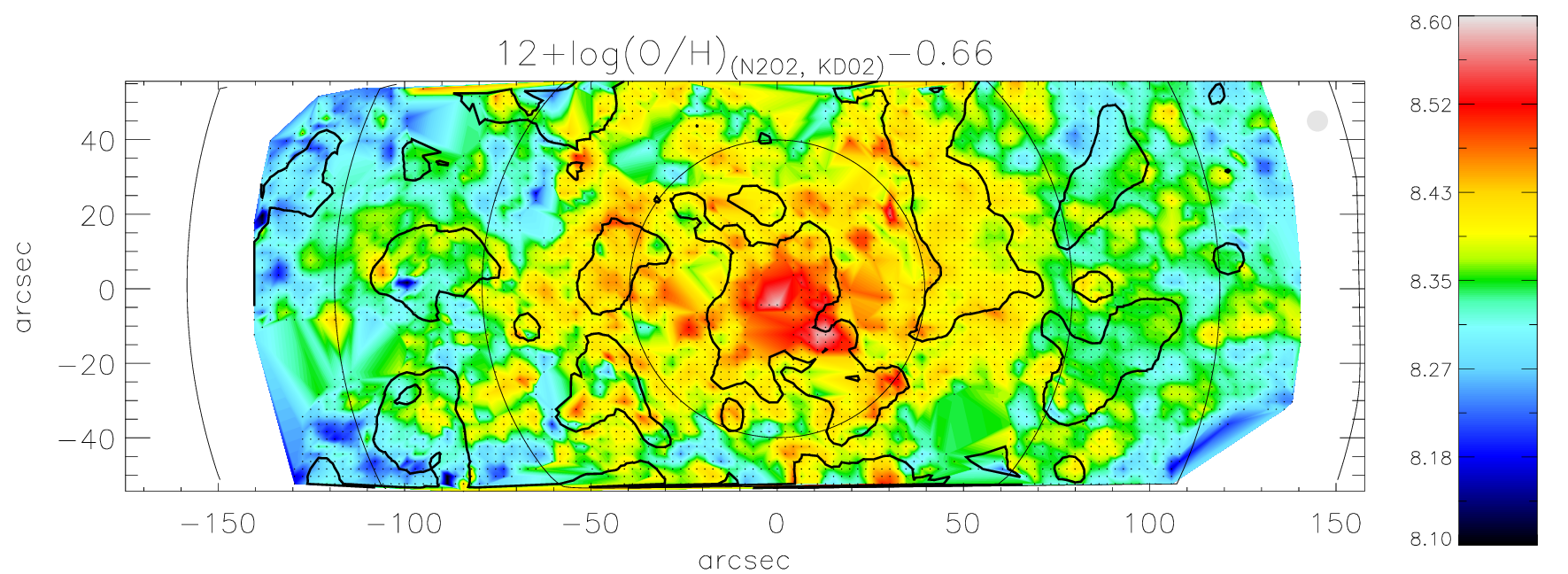}
\plotone{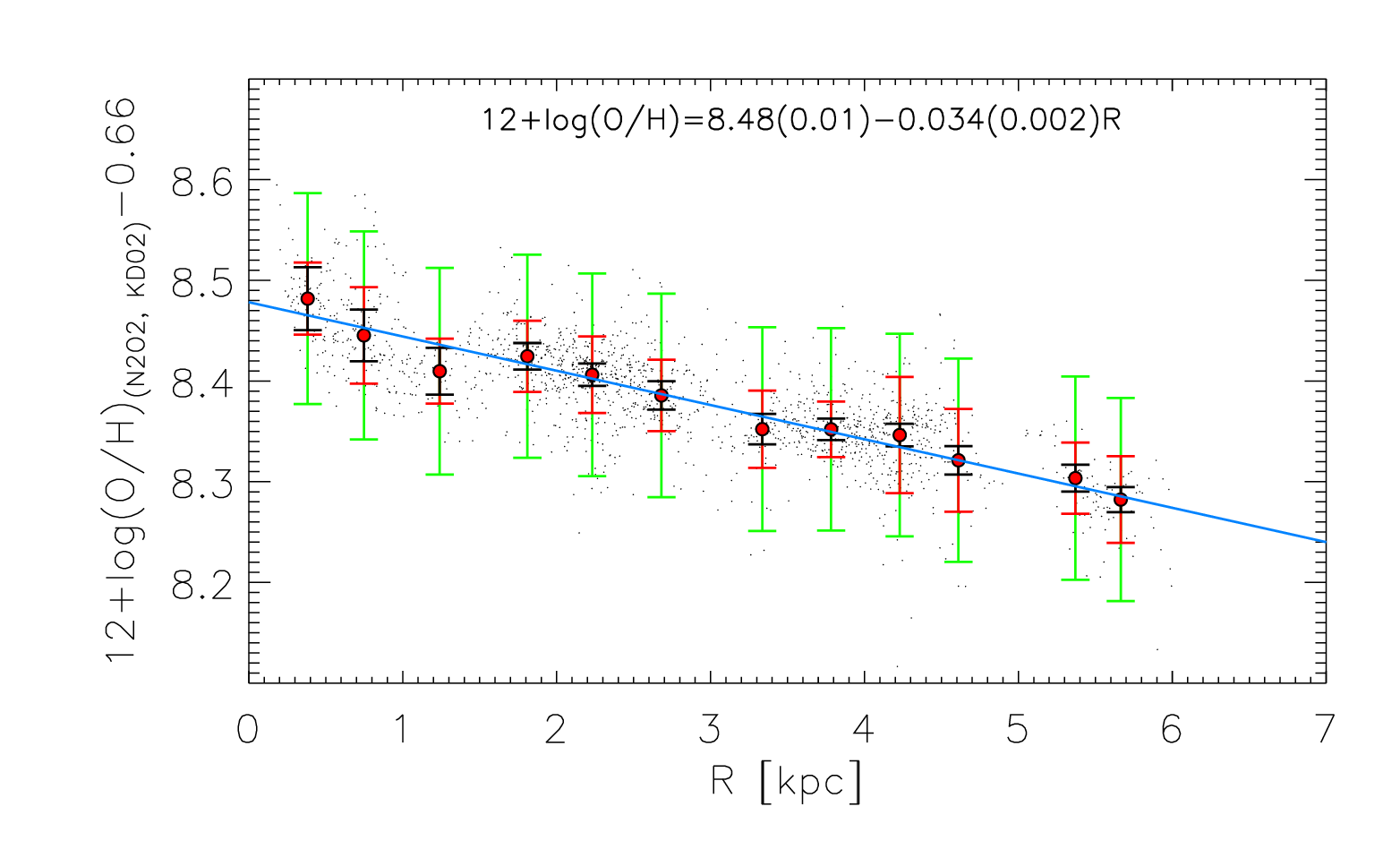}
\caption{Spatial $N2O2$ metallicity distribution in NGC 628 measured
  from the VENGA data-cube, offset to
  match the ``empirical method'' metallicity scale. The top panel
  shows a two-dimensional map of 12+log(O/H).
  Thick contours marks regions above the H$\alpha$ surface
brightness cut where the nebular emission is dominated by HII
regions. The bottom panel shows the radial distribution in metallicity
of all spaxels inside these regions (black dots). Red circles and
error-bars mark the median and standard deviation of all spaxels in
0.5 kpc radial bins. Black error-bars show the median measurement error in each
radial bin, and green error-bars show the 0.1 dex intrinsic scatter
associated with the $N2O2$ calibration}
\label{fig-4}
\end{center}
\end{figure}
 
On the other hand, the metallicity compilations for the systems with
measured $X_{\rm CO}$ values reported in \cite{bolatto08} and
\cite{leroy11} are largely based on direct electron temperature
measurements. The metallicities reported in \cite{schruba12} are on an
intermediate scale, since the authors averaged the values from the two
indicators used by \cite{moustakas10}: the photo-ionization
model calibrated $R_{23}$ metallicities of \cite{kobulnicky04}
(hereafter $R_{23}-KK04$) and the
empirically calibrated $R_{23}$ method of \cite{pilyugin05} (hereafter
$R_{23}-PT05$). 

We decide to apply a simple offset in metallicity to both our $N2O2$
measured values in NGC 628 and the reported metallicity values in
\cite{schruba12}, to put them on a common scale with the direct method
metallicities reported in \cite{bolatto08} and \cite{leroy11}. To do
so, we measured the metallicity in the NGC 628 data-cube
using the three methods mentioned above ($N2O2$, $R_{23}-KK04$, and
$R_{23}-PT05$), and measure the mean offset in the recovered
metallicity values between different methods for all spaxels above a H$\alpha$ surface
brightness cut chosen to trace HII regions \footnote{All these
  strong-line methods have been calibrated either empirically against
  direct-method metallicities of HII regions, or photo-ionization
  models with densities and ionization parameters typical of HII
  regions, therefore they cannot be applied to low surface brightness
  regions in which the nebular emission is dominated by the diffuse
  ionized gas.}. As expected, both the $N2O2$ and the average between
the $R_{23}-KK04$, and $R_{23}-PT05$ methods (i.e. the method used in
\cite{schruba12}) yield higher metallicities than $R_{23}-PT05$ alone,
which is calibrated against direct method measurements on HII
regions and should fall in a similar scale to the values reported in
\cite{bolatto08} and \cite{leroy11}. We find and apply mean offsets of
0.66 dex and 0.35 dex to the $N2O2$ and \cite{schruba12} metallicities
before doing any comparisons. These values are in good agreement with
the 0.6 dex offset found between $R_{23}-KK04$, and $R_{23}-PT05$ by
\cite{moustakas10}.

Note that we decided to apply a single offset in metallicity instead of using the
conversion relations of \cite{kewley08}. This is done deliberately, as
the non-linear nature of these conversions imply that the measured
metallicity gradient would change its value depending on the adopted
scale. Also, these conversions have been calibrated against integrated
measurements of galaxies, and it is not clear how valid it would be to apply
them to spatially resolved regions within a single object.
This issue will be the subject of a future publication in which
we will analyze the impact of different methods on the observed
metallicity distributions in the VENGA galaxies. For the purpose of
this work, we prefer to conserve the shape of the gradient as measured
by the $N2O2$ method which we consider more robust than other
formulations. In any case, using the \cite{kewley08} conversions, or
using $R_{23}-PT05$ directly on our data instead of $N2O2$ does not
significantly change our results.

Figure \ref{fig-4} presents the spatial distribution in metallicity of
NGC 628. The top panel shows a two-dimensional map of 12+log(O/H),
where thick contours marks regions above the H$\alpha$ surface
brightness cut where the nebular emission is dominated by HII
regions. The bottom panel shows the distribution of metallicity as a
function of galactocentric radius for data-cube spaxels inside these
regions. Black error-bars show the median measurement errors in each
radial bin, red error-bars mark the standard deviation for all spaxels
in each bin, and green error-bars show the 0.1 dex intrinsic scatter
associated with the $N2O2$ calibration \citep{kewley08}. We measure a
gradient of \mbox{$\Delta{\rm log}(OH)=0.036\pm0.002$ dex kpc$^{-1}$},
in good agreement with the IFU measurements of \cite{rosales-ortega11}.

\begin{figure*}
\begin{center}
\epsscale{1.2}
\plotone{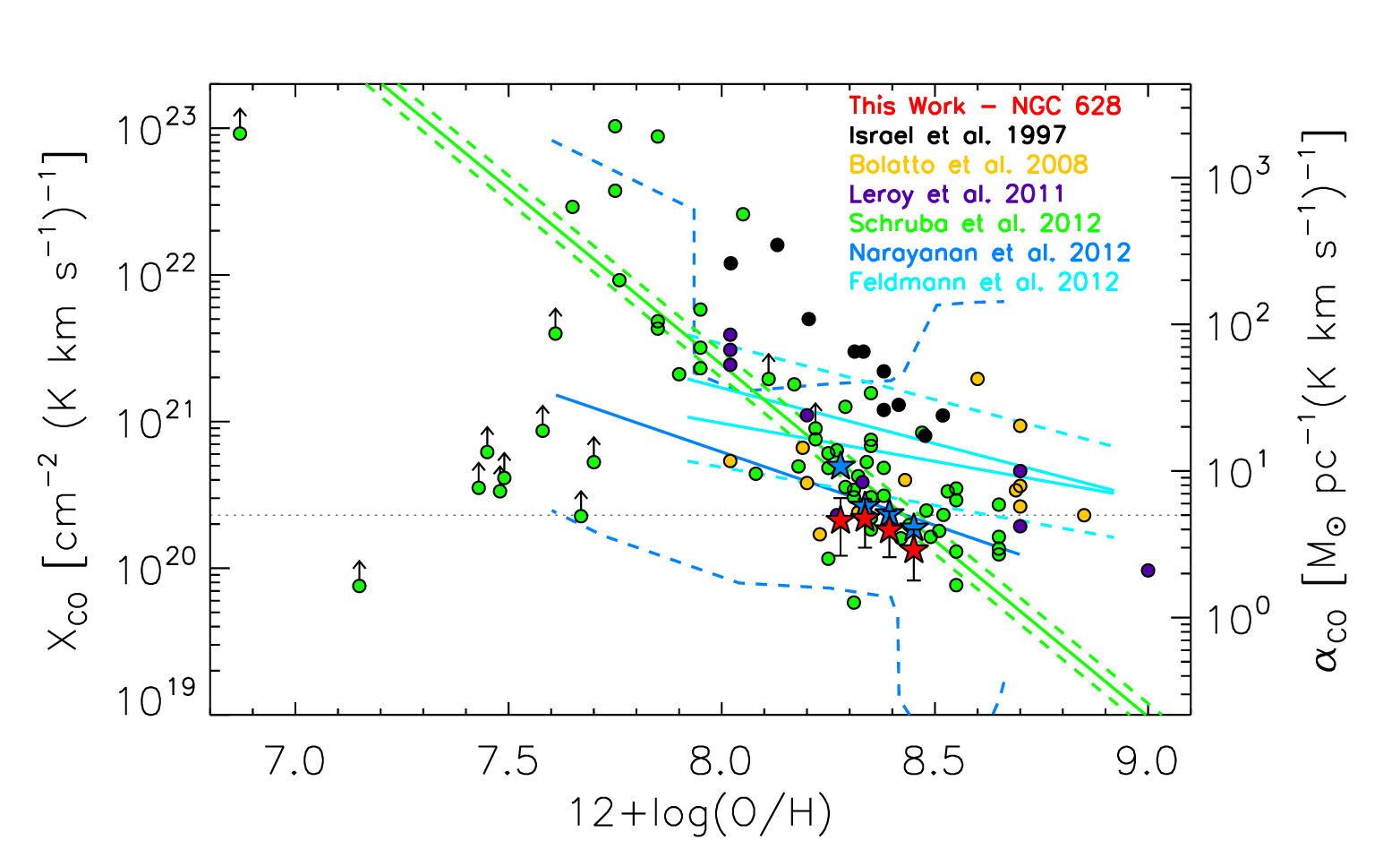}
\caption{$X_{\rm CO}$ factor as a function of metallicity
  (12+log(O/H)). Red stars correspond to the four radial bins in
  NGC 628 analyzed in this work. The measurements of \cite{bolatto08},
  \cite{leroy11}, and \cite{schruba12} are shown as orange, purple,
  and green circles. The solid and dashed green lines show the best-fit
relation to the clean HERACLES sample in \cite{schruba12}. The
theoretical models of \cite{narayanan12} and \cite{feldmann12} are
also shown. For the \cite{narayanan12} model, the solid blue line show
the expected scaling of $X_{\rm CO}$ with metallicity for a constant
$\Sigma_{H2}=100$ M$_{\odot}$pc$^{-2}$, and the blue dashed lines mark
the scatter in $X_{\rm CO}$ seen across their simulated galaxies. Blue
stars mark the predictions from the model taking into account the
differences in gas density for each radial bin in NGC 628. For the
\cite{feldmann12} model the two solid cyan lines mark, from top to
bottom, their ``constant line-width'' and ``virialized'' models at 4
kpc resolution. Dashed lines mark factors of 2 above and below the two
models respectively, which the authors state bound the expected
scatter in $X_{\rm CO}$ for their models.}

\label{fig-5}
\end{center}
\end{figure*}

\subsection{$X_{\rm CO}$ vs. Metallicity}

In Figure \ref{fig-5}, red stars show the $X_{\rm CO}$ values measured in each radial
annuli in NGC 628, as a function of the metallicity at the radius corresponding
to the center of each annuli (evaluated from the metallicity gradient
measured in \S5.1). Also shown are the data of \cite{bolatto08},
\cite{leroy11}, and \cite{schruba12}\footnote{The data-points of
  \cite{schruba12} have been scaled up by a factor of 1.11 to
  account for the difference in the assumed depletion timescale (2.0 vs 1.8 Gyr)}, as well as predictions from the
theoretical models of \cite{narayanan12}
and \cite{feldmann12}.

Across the disk of NGC 628, $X_{\rm CO}$ increases towards lower
metallicity regions. This is expected from the positive and negative gradients
measured for $X_{\rm CO}$ and the oxygen abundance respectively (\S4 and \S5.1).
This behavior is consistent with integrated measurements of galaxies
in the local universe. Both \cite{leroy11} and \cite{schruba12} see an
increase in $X_{\rm CO}$ towards low metallicities by conducting
integrated measurements across galaxies using dust mass modeling in
the first case, and a similar method to the one used here in the
later. The $X_{\rm CO}$ values inferred from GMC virial masses in
\cite{bolatto08} do not show an obvious trend
with metallicity, but are consistent with other studies given the
large scatter in the observed relation. Also, as discussed in the
original \cite{bolatto08} paper, and in
\cite{schruba12}, estimates of $X_{\rm CO}$ using virial mass measurements
from CO observations most likely miss the ``CO dark'' envelopes of
molecular clouds in low metallicity galaxies, as they can only trace
mass enclosed in the CO emitting region. This could translate in an
underestimation of $X_{\rm CO}$ at low metallicity in the \cite{bolatto08}
data. The same effect could explain the constancy of $X_{\rm CO}$ seen by
\cite{rosolowsky03} across 45 GMCs in M33, even in the
presence of a 0.8 dex change in metallicity within their sample. 

The NGC 628 datapoints are consistent with previous
observations, specially considering the large scatter seen in the
$X_{\rm CO}$ metallicity relation ($\sim0.5$ dex). The limited dynamic
range in metallicity that we can sample across the inner 7 kpc of a
single galaxy prevents us from being able to put strong constraints in
the slope of the $X_{\rm CO}$-metallicity relation. Modeling this
relation as a power-law of the form $X_{\rm CO}\propto Z^\alpha$, \cite{schruba12}
measures a slope of $\alpha=-2.0\pm0.4$ for the full HERACLES sample after rejecting
starburst galaxies (green solid line in Figure \ref{fig-5}), while
\cite{israel97} finds a slope of $\alpha=-2.7\pm0.3$ in their study of dust
emission in local dwarfs. A linear fit to our data points yields a
slope of $\alpha=-1.1\pm1.6$, shallower than previous measurements but still
consistent with both \cite{israel97} and \cite{schruba12}. Again, given the limited dynamic
range in metallicity of our data, and the large scatter in the
$X_{\rm CO}$-metallicity relation, detailed comparisons are difficult. 

The slope of the observed relation in NGC 628 is also in agreement with expectations from
theoretical models. At fixed molecular gas surface density, the
simulations in \cite{narayanan12} predict a linear dependance of
$X_{\rm CO}$ with metallicity (i.e. $\alpha=-1.0$)\footnote{Note that in
  their model this exponent changes if one asumes constant CO brightness instead of
constant $\Sigma_{H2}$ because of the metallicity dependance of
$X_{\rm CO}$ (see Equations 6 and 8 in \cite{narayanan12})}, and the study by
\cite{feldmann12} predicts slopes in the  $-0.5$ to $-0.8$ range,
depending on the treatment given to the sub-grid gas dynamics in their
simulations. Regarding the normalization of the observed relation, our data
shows a remarkably good agreement with the \cite{narayanan12} model
for an assumed molecular surface density of $\Sigma_{H2}=100$
M$_{\odot}$pc$^{-2}$. On the other hand, the \cite{feldmann12} model,$\alpha=-2.7\pm0.3$
in which $X_{\rm CO}$ is independent of the molecular gas surface density,
predicts values that are a factor of $\sim4$ higher than our
observations. In the following section we discuss the characteristics
of these models in more detail, paying particular attention to the
discrepancies which arise from the different treatments given to the role
of the local gas surface density at setting the $X_{\rm CO}$ conversion factor.\\

\section{The Role of Gas Density and the
  UV Radiation Field at setting $X_{\rm CO}$}

In the previous section we explored the role of metallicity in
setting the value of $X_{\rm CO}$, and discussed how well observations are
described by the dependance on metallicity predicted by current
theoretical models of CO emission in molecular gas. Here, we explore
the dependance of $X_{\rm CO}$ on other physical properties like the
surface density of molecular gas, and the local UV radiation field, which at
the same time, affect the temperature and degree of turbulence in the
ISM. Informed by the predictions of theoretical models, we examine
these potential trends in the NGC 628 data, and explore if other
parameters, beyond the metallicity, are important at setting the
observed $X_{\rm CO}$ radial profile.

\subsection{Further Comparison to Theory: Gas Density}

In the theoretical model of \cite{feldmann12}, $X_{\rm CO}$ is a strong function of the molecular
gas surface density of a given GMC, with the conversion factor growing
away from its minimum towards both low and high densities. This behavior is a
consequence of two different effects taking place in the low and high
density regimes. At high column densities the numerator in Equation 1
increases linearly, while the denominator (i.e. the CO intensity) is
either constant or increases sub-linearly (as in the virialized cloud
approximation for which $I(CO)\propto \Delta v \propto
\sqrt{N(H_2)}$). This effect is supported by spatially resolved observations of GMCs in
the Milky Way \citep{heiderman10}. Therefore $X_{\rm CO}$ increases towards high
column densities. On the other hand, towards lower $N(H2)$, the CO
abundance drops due to CO dissociation as the dust extinction (which
is proportional to the column density) decreases, giving rise to large
``CO dark'' envelopes and also implying an increase of $X_{\rm
  CO}$. Both the location of the minimum, and the value of $X_{\rm CO}$ at the minimum
are strong functions of the metallicity in the model. 

When attempting to study these trends on kpc scales within or
across galaxies, we must take into account the fact that we measure
averages for ensembles of molecular clouds with a distribution in their
physical properties. In the \cite{feldmann12} model, this average
washes away the dependance on column density seen on small
scales, and the authors predict almost no variation in $X_{\rm CO}$
as a function of H$_2$ surface density. It is important to point
out that in this model the molecular gas
temperature is fixed to 10 K in the calculation of $I(CO)$, and the
authors either impose a constant CO line-width, or assume virialized
clouds (i.e. $\Delta v \propto \sqrt{N(H_2)}$).

As discussed in the previous section, the models of \cite{narayanan12}
and \cite{feldmann12} predict similar behaviors for $X_{\rm CO}$ as a
function of metallicity. On the other hand, the simulations conducted
by \cite{narayanan12} do not assume a fixed temperature for the
molecular component of the ISM, and instead include relevant heating and
cooling calculations, which allow the authors to trace the molecular
gas temperature. Furthermore, for molecular clouds which are larger
than the spatial resolution of the simulation, they use the simulations
themselves to follow the dynamical state of the molecular gas (i.e. the
gas velocity dispersion sets the CO line-width). For
sub-resolution clouds they assume virialization.

The ability to follow the temperature and dynamical state of the gas,
allows \cite{narayanan12} to study the impact of turbulence and
heating from both ongoing star formation, and large scale gas dynamics, on
the $X_{\rm CO}$ factor under high density conditions like the ones present in starbursts,
mergers, and the central regions of galaxies.
While gas density itself should have little direct impact on
$X_{\rm CO}$ when averaging over kpc scales, the existence of a
correlation like the SFL implies that higher molecular gas density
is accompanied by stronger star formation activity. Star formation is
associated with feedback from proto-stellar jets, stellar
winds, and supernovae (SN) explosions, as well as heating (both
photoelectric and by cosmic rays produced in SN), which rises both the
temperature and turbulence of the gas. Furthermore, high density
regions are also commonly associated with dynamically violent
environments, in which for example, merger induced shocks and
cloud-cloud collisions can induce higher levels of turbulence in the gas.
Given the optically thick nature of the CO(1-0) transition, any
turbulence induced line broadening translates into a brighter $I(CO)$ and a lower
$X_{\rm CO}$. Similarly, a higher gas temperature translates into a higher
brightness temperature for the CO line. By construction, these effects
are not recovered in the \cite{feldmann12} simulations.

Comparing the observed dependance of $X_{\rm CO}$ with gas surface density, or with
CO surface brightness, to theoretical numerical model predictions is
challenging, and a series of caveats arise which must be kept in mind
when interpreting this type of comparison. These caveats arise on one
side from the limitations of the observations, where the large physical spatial
resolution typically achieved in extragalactic studies ($\sim600$ pc
in NGC 628 for the HERACLES beam-size) dilutes the actual surface
brightness (or mass surface density) of GMCs in the beam, by some
factor which depends on both the intrinsic surface brightness
distribution of the clouds, and the covering fraction of clouds across the
area over which the emission is being integrated. In denser regions,
overlap of optically thick clouds can also affect the measured surface
brightness. On the other hand, in numerical simulations, while
physical numerical quantities are readily accesible, the definition of
a ``cloud'', and the exact method used to measure its
surface brightness or density can be non-trivial. This can be
especially problematic when the resolution of the simulation is larger,
or of the same order of magnitude as the typical sizes of clouds.

\cite{narayanan12} define $X_{\rm CO}$ as the ratio between the average
column density, and the average CO intensity (luminosity divided by
area) for each of their simulated
galaxies, and provide a fitted formula to calculate $X_{\rm CO}$ from
measurements of the metallicity ($Z'=Z/Z_{\odot}$) and the luminosity-weighted CO
intensity of GMCs ($\langle I(CO)_{GMC} \rangle$) in a galaxy:

\begin{equation}
X_{\rm CO}=\frac{{\rm min}[4,\;6.75\times \langle I(CO)_{GMC} \rangle ^{-0.32}]\times10^{20}}{Z'^{0.65}}
\end{equation}

Unfortunately, $\langle I(CO)_{GMC} \rangle$ is not
recoverable by CO observations at kpc scales, and also note that it is not the same
quantity used in their definition of $X_{\rm CO}$. Instead, of the
luminosity-weighted average of the CO intensity of individual clouds, we
measure the average CO intensity over large areas of the galactic disk of NGC
628, which is equivalent to the average intensity going in the definition
of $X_{\rm CO}$ in the model.

Because of averaging over large areas, with a lower than unity cloud
filling factor, our measured $I(CO)$ values should be significantly lower than
the CO intensities one would measure for individual GMCs. In order
to use the \cite{narayanan12} model to make meaningful predictions
from our data, we adopt a ``clumping factor'' to scale our observed average CO
intensities, in an attempt to recover the luminosity weighted average intensity
of individual GMC in the observed regions. A similar approach is adopted
by \cite{krumholz09} to transform observed average H$_2$ surface
densities into intrinsic GMC surface densities for input to their SFL
model. 

The choice of the clumping factor is somewhat arbitrary, but it only
affects the absolute value of the predicted $X_{\rm CO}$ factors, and not
any relative trends observed across the galaxy. We use
a factor of 30 to scale up the observed $I(CO)$ values for input into
Equation 10 in \cite{narayanan12}. The
justification for adopting this value is as follows. For a MW
$X_{\rm CO}$ factor (which is consistent with the average factor we
measure in NGC 628), the average CO intensity of NGC 628 inside a 7 kpc radius
implies a mean value of $\Sigma_{H2}$ which 9 times lower than the typical surface
density of  $\Sigma_{H2}\simeq100$ M$_{\odot}$pc$^{-2}$ measured for GMCs in
the Milky Way and the Local Group \citep{roman-duval10, rosolowsky03,
  bolatto08}. On the other hand, comparison of Equations 6 and 10 in
\cite{narayanan12} implies that in their model $\Sigma_{H2}
\propto \langle I(CO)_{GMC} \rangle^{0.64}$. Therefore, a clumping factor
of $30\simeq9^{1/0.64}$ would imply that the luminosity weighted
average GMC surface density in the inner 7 kpc of NGC 628 is similar to the
typical MW and Local Group value.

The predicted $X_{\rm CO}$ values for different radial bins in NGC 628, computed using the measured
metallicity and observed CO intensity (average between the HERACLES,
BIMA, and CARMA measurements) corrected by this clumping factor,
are shown as blue stars in Figure \ref{fig-5}. Predictions for the
\cite{feldmann12} model would fall right on top of one of the two cyan solid
lines in Figure \ref{fig-5}, as in their parametrization $X_{\rm CO}$ is
independent of $\Sigma_{H2}$ and $I(CO)$. As stated in \S5, the
\cite{narayanan12} model is in better agreement with our data than
the model of \cite{feldmann12}.

\begin{figure}[t]
\begin{center}
\epsscale{1.2}
\plotone{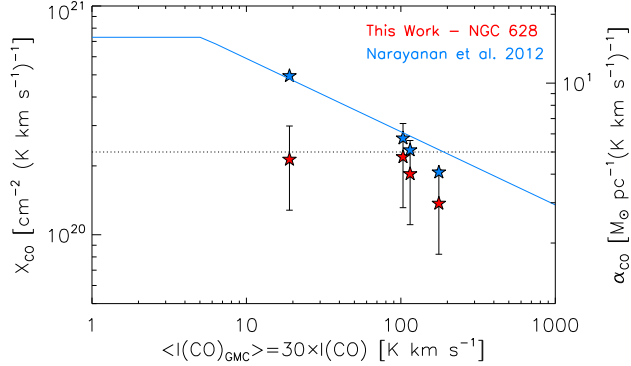}
\caption{$X_{\rm CO}$ as a function of CO surface brightness for the four
  radial bins studied in NGC 628. CO intensities have been scaled by a
  clumpiness parameter of 30. Red stars show our measurements and
  blue stars show the predictions from the \cite{narayanan12} model
  using the metallicity and CO intensity of each bin. The solid blue
  line shows the predictions from the model at constant metallicity
  (the characteristic metallicity at 0.4$R_{25}$). The horizontal
  dotted line shows the canonical MW value from \cite{pineda10}.}
\label{fig-6}
\end{center}
\end{figure}

Figure \ref{fig-6} presents a comparison between the average
$X_{\rm CO}$ values for each radial bin in NGC 628 (red stars) and the values predicted by the
\cite{narayanan12} model (blue stars). Both are presented as a function
of $I(CO)$ scaled by a clumpiness factor of 30, which, as stated above,
should be roughly equivalent to the input $\langle I(CO)_{GMC} \rangle$ for the
model. By inspecting Figures \ref{fig-5} and \ref{fig-6}  we can see that the model reproduces
very well the observed trends in $X_{\rm CO}$ with both metallicity and
CO surface brightness for the inner three radial bins in NGC 628. The
slopes of both trends are consistent, and the absolute
values agree to 1$\sigma$. 

A caveat to the above approach is that adopting a single clumpiness
parameter is equivalent to assuming a constant cloud filling
factor over the observed area. A decrease in the cloud filling factor
towards the outer disk is in fact likely, and could indeed explain
both the flattening seen in $X_{\rm CO}$ for the $N=1$ case when using the
stacked \cite{schruba11} data in the outer disk of NGC 628, and the
discrepancy observed between the data and the \cite{narayanan12} model
for the outer radian bin. This latter discrepancy is reduced if one assumes
$N=1.5$ instead of $N=1$ for the SFL.

At this point, it is important to remind the reader about the many
scaling factors which affect the normalization of the $X_{\rm CO}$
values. These include the zero-points in the flux calibration of the
HERACLES CO and VENGA H$\alpha$ maps, and the adopted H$\alpha$ SFR calibration
(we have scaled the other datasets to this reference), the adopted
offsets used to match the metallicity scale of the different datasets
and the models, and the adopted clumpiness parameter which is
constrained only by demanding a typical GMC surface density of
$\Sigma_{H2}\simeq100$ M$_{\odot}$pc$^{-2}$. Given the
inherent systematic uncertainties associated with all these scaling
factors, it is difficult to draw strong conclusions from the relative
offset seen between the observed and predicted $X_{\rm CO}$
values. None of these scaling factors affect the relative
trends seen across different radial bins in both the data and the model.

The main conclusion we can draw from the above comparison is the
following. Our observations are in good agreement with the
\cite{narayanan12} model. If $X_{\rm CO}$ follows the metallicity
and CO brightness dependance proposed in this model, that is
$X_{\rm CO}\propto \langle I(CO)_{GMC} \rangle ^{-0.32}\times Z'^{-0.65}$, then, the
observed metallicity and CO surface brightness gradients imply that
both quantities contribute to the formation
of an $X_{\rm CO}$ gradient across the disk of NGC 628. As stated in \S5,
the measured metallicity gradient implies $Z'\propto 10^{-0.04R}$, and a
linear fit to the CO surface brightness of the inner three radial bins
($R<5$ kpc) implies $\langle I(CO)_{GMC} \rangle \propto 10^{-0.07R}$. The model implies that $X_{\rm CO}
\propto 10^{(0.32\times0.07+0.65\times0.04)R}$, or a gradient of \mbox{$\Delta
{\rm log}(X_{\rm CO})=0.05$ dex kpc$^{-1}$}, in
excellent agreement with the measured gradient of  \mbox{$\Delta
{\rm log}(X_{\rm CO})=0.06\pm0.02$ dex kpc$^{-1}$}. 
From the above calculation it can be seen that
both the metallicity and the CO surface brightness (or equivalently
the molecular gas surface density) contribute to the observed
$X_{\rm CO}$ gradient. The \cite{narayanan12} prescription, in which
$X_{\rm CO}$ is inversely proportional to the metallicity, and inversely
proportional to the square root of the molecular gas surface density,
is consistent with our data.

Our results indicate that across the disk of NGC 628, both the metallicity and the
molecular gas surface density (through its incidence in the
temperature and turbulence of the gas) are
important factors setting the CO to H$_2$ conversion factor.
Future analysis of a larger subset of the VENGA sample will allows us
to know if this result is general enough to be applicable to most
massive spiral galaxies in the local universe.

\subsection{$X_{\rm CO}$ and the UV Radiation Field}

In this section we use the ionization parameter, as measured from HII
region emission line ratios in the VENGA data, as a proxy for the
intensity of the local interstellar UV radiation field $S_{UV}$ across the disk
of NGC 628. The goal is to study the potential effect that the UV
radiation field has at setting the value of $X_{\rm CO}$.

The strength of the local UV radiation field has been proposed as an
important factor setting the CO to H$_2$ conversion
factor. Analyzing a sample of individual molecular complexes in the LMC and the
SMC, \cite{israel97} finds a linear correlation between the
$X_{\rm CO}$ and the ratio of the far-IR surface brightness to HI column
density ($\sigma_{FIR}/N(HI)$) which the author considers a proxy for
the local UV radiation field per hydrogen atom (reprocessed after
absorption and reemission by dust). 

The effect of the UV radiation field on $X_{\rm CO}$ is also discussed by
\cite{feldmann12} in the context of their numerical simulations and
their model of CO emission. The authors propose that for a single
molecular cloud, the CO abundance decreases in the presence of a
stronger UV radiation field $U_{UV}$ (which they parametrize in units
of the local interstellar UV radiation field, so
\mbox{$U_{UV}=S_{1000\AA}/S_{MW}$},  with \mbox{$S_{MW}=10^6$ photons cm$^{-2}$
s$^{-1}$ eV$^{}-1$).} This dependance is stronger at
low $A_V$ (i.e. low $N(H_2)$, or low $Z$, or both), but an order of
magnitude change in $U_{UV}$ can still change the CO abundance by factors of a few at
$A_V=1-3$ mag. These extinction levels correspond to the outer
envelopes CO emitting regions in GMCs, where CO(1-0) is
starting to become optically thick. As the UV radiation field goes up,
the CO abundance goes down, and $X_{\rm CO}$ increases for a single
cloud (see their Figure 2).  

When investigating this effect on galactic scales, \cite{feldmann12}
finds that the strong dependance of $X_{\rm CO}$
with the UV radiation field present in their model for single GMCs at fixed $N(H_2)$,
is completely suppressed in their simulated
galaxies. This is due to the fact that in the simulations, physical
parameters are measured over ensembles of molecular clouds with
a distribution in their properties, and the abundance of clouds with
low $A_V$, and therefore high $X_{\rm CO}$, decreases strongly for higher
values of the UV radiation field. In the \cite{feldmann12}
simulations, increasing $U_{UV}$ suppresses the low density
tail of the $N(H2)$ distribution function, dissociating diffuse molecular
clouds with low densities ($\lesssim 10^{21}$ cm$^{-2}$) which have
large $X_{\rm CO}$ values. Therefore, when averaging over kpc
scales, the radiation field has little impact in the average $X_{\rm CO}$
value, which is dominated by the surviving higher density regions with
$N(H_2)\simeq10^{22}$ cm$^{-2}$.

\begin{figure}
\begin{center}
\epsscale{1.2}
\plotone{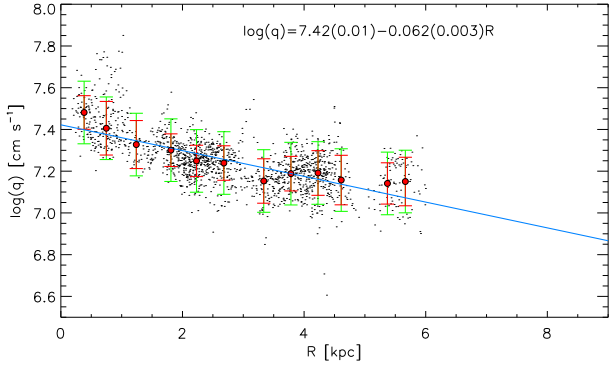}
\caption{Ionization parameter as a function of radius in NGC
  628. Black dots correspond to spaxels dominated by HII region
  emission in the VENGA data-cube. Red circles show the median values in 0.5 kpc wide radial bins, and
red and green error bars show the standard deviation in each bin and
the systematic uncertainty in the calibration respectively.}
\label{fig-7}
\end{center}
\end{figure}

A competing effect, which might be important at setting the value of
$X_{\rm CO}$, is photoelectric heating working on large polycyclic
aromatic hydrocarbon (PAH) molecules and small dust grains in
the outer envelopes of molecular clouds \citep{tielens05, glover10}. 
The photoelectric heating
rate is directly proportional to the intensity of the interstellar UV
radiation field \citep{tielens05}, therefore, the gas temperature, and
the CO transition brightness temperature can be increased in the
presence of a stronger UV radiation field. By construction, this
effect cannot be recovered by \cite{feldmann12} as
the authors impose a constant molecular gas temperature of 10K in
their simulations. This effect might be present in the
\cite{narayanan12} simulations, which follow the main heating and
cooling processes in the gas, and are able to trace changes in the gas
temperature, though the authors do not attempt to separate it from
dynamical effects on the brightness of the CO line. As mentioned
above, part of the dependance of $X_{\rm CO}$ with molecular gas surface
density might indeed be associated with this mechanism, as higher gas
density translates into higher star formation activity and hence, a
stronger interstellar UV radiation field.

In Figure \ref{fig-7} we present the radial distribution of the
ionization parameter 

\begin{equation}
q=\frac{S_{H^0}}{n}
\end{equation}

\noindent
where $S_{H^0}$ is the ionizing photon flux per unit area, and $n$ is
the number density of hydrogen atoms. We have estimated $q$ from the
[OIII]$\lambda$5007/[OII]$\lambda$3727 line ratio, as measured from
the VENGA data-cube of NGC 628, following the iterative
procedures described in \cite{kobulnicky04}. The measurements
are limited to spaxels dominated by HII region emission (black dots) as in the case
of the metallicity measurements described in \S5.1. Red circles in
Figure \ref{fig-7} show median values in at low
densities ($\lesssim 10^{21}$ cm$^{-2}$) 0.5 kpc wide radial bins, and
red and green error bars show the standard deviation in each bin and
the systematic uncertainty in the calibration respectively. 

In NGC 628, the ionization parameter falls towards larger
radii. A linear fit implies a gradient of $\Delta {\rm
  log}(q)=0.061\pm0.003$ cm s$^{-1}$ kpc$^{-1}$. A drop in the
ionizing photon flux per hydrogen atom is consistent with the results
of \cite{aniano12} who find a decreasing level of dust heating with
radius due to a decrease in starlight intensity by modeling the {\it
  Spitzer}+{\it Herschel} spatially resolved dust SED
of NGC 628. Considering
the $X_{\rm CO}$ radial profile measured in \S4, we find no evidence of an
increase in $X_{\rm CO}$ in regions where the UV radiation field is
enhanced, but we rather observe the opposite trend. 
The increase in $X_{\rm CO}$ for higher $U_{UV}$
expected on clouds scales in the model of \cite{feldmann12}, and
observed on molecular complexes scales by \cite{israel97}, is not seen
in kpc scales across the disk of NGC 628. This could be cause in part
by the effects of averaging over large ensembles of clouds, discussed
in \cite{feldmann12} and described above, but this could only suppress
this dependance. 

The fact that we observe the opposite behavior, is consistent with the
photoelectric heating effect described above. It is difficult to
decouple the impact that the UV radiation field might have on the gas
temperature and hence, on the value of $X_{\rm CO}$, from surface density
dependance discussed \S6.1. All we can conclude, is that the observed
radial distribution in the ionization parameter, implies that thermal
effects might be important at setting $X_{\rm CO}$. Further study will be
necessary to properly model these thermal effects, and decouple them
from dynamical effects which also impact the radiative transfer of CO
photons.

\section{Conclusions}

By inverting the SFL we obtain an independent estimate of the H$_2$
surface density across the disk of NGC 628 from the measured SFR
surface density. Comparison to the observed CO intensity yields a
measurement of the $^{12}$CO(1-0) to H$_2$ conversion
factor ($X_{\rm CO}$). By studying the radial profile of $X_{\rm CO}$, and its
relation to other quantities like the metallicity, CO surface
brightness, and the ionization parameter across the disk of the galaxy
we reach the following conclusions:

\begin{itemize}

\item The $X_{\rm CO}$ factor increases as a function of radius across the disk
of \mbox{NGC 628}. A linear fit to the data in the inner 7 kpc of the
galaxy implies a gradient of  $\Delta {\rm
    log}(X_{\rm CO})=0.06\pm0.02$ dex kpc$^{-1}$ under the assumption of
a $N=1$ SFL slope. The radial profile is in agreement with
measurements using an independent technique based on dust emission modeling
and simultaneous fitting of the dust-to-gas ratio and the $X_{\rm CO}$
conversion (Karin Sandstrom private communication).

\item The observed $X_{\rm CO}$ radial profile is independent of the
  adopted SFR tracer (H$\alpha$ vs FUV+24$\mu$m), and the CO emission
  map used (single-dish vs. interferometer, CO(1-0) vs CO(2-1)).

\item The observed $X_{\rm CO}$ radial profile is robust against changes
  in the adopted slope for the SFL ($N=1$ vs 1.5). While assuming a
  steeper slope slightly steepens the observed profile, the effect is
  small compared to the changes seen in $X_{\rm CO}$ as a function of
  radius in NGC 628. The observed radial profile is robust, and it is
  not strongly affected by potential changes in the shape of the
  assumed SFL across the galaxy.

\item The observed metallicity gradient in NGC 628 implies that
  regions of lower metallicity show larger $X_{\rm CO}$ values. This is in
  agreement with integrated measurements across samples of galaxies in
  the local universe, and the predictions of theoretical models of CO
  emission in molecular gas.

\item Regions of lower CO surface brightness show higher $X_{\rm CO}$
  values. This is in agreement with theoretical models which predict
  an enhanced escape probability of CO photons in higher density
  regions, due to an enhanced star formation activity which translates
  in a broadening of the CO line due to elevated gas temperatures and
  turbulence.

\item Informed by the theoretical model of \cite{narayanan12} we
  conclude that both the dependances with metallicity and H$_2$ surface
  density contribute in roughly similar amounts to the formation of
  the observed gradient in $X_{\rm CO}$.

\item The ionization parameter shows a linear decreasing gradient as a
  function of radius, so regions where the intensity of the UV
  radiation field is higher show lower values of $X_{CO}$. This
  suggest that photoelectric heating might have an impact at setting
  the brightness temperature of the CO line, and therefore the value
  of the CO to H$_2$ conversion factor.

\item When comparing with theoretical models, our observations agree
  very well with the predictions of \cite{narayanan12}. While our
  data also agrees with the model of \cite{feldmann12} in the slope of
  the $X_{\rm CO}$-metallicity relation, and the lack of dependance of
  $X_{\rm CO}$ with the local ionizing field on kpc scales, this model
  predicts $X_{\rm CO}$ values which are typically a factor of $\sim 4$
  higher than the observed ones.

\item Given the observed $X_{\rm CO}$ radial profile in NGC 628, we
  conclude that using a single MW $X_{\rm CO}$ factor to estimate the total H$_2$
  mass would imply an overestimation of 20\% in this quantity. While
  this is a relatively small effect, our results imply that much
  larger systematic deviations can occur when using a canonical value
  in specific regions within galaxies, or when the local conditions of
  the ISM in the regions of interest differ significantly from the
  canonical values.

\end{itemize}

A series of assumptions have been made in order to reach these
conclusions, and it is important to keep these in mind while
interpreting our results. The uncertainty in the flux calibration and
sensitivity to emission on different spatial scales in the three CO
datasets used, the use of a constant non radially dependent
CO(2-1) to CO(1-0) ratio to scale the HERACLES data, the uncertainty
in extinction correction (in the case of H$\alpha$) and the SFR
calibrations used for H$\alpha$ and FUV plus 24$\mu$m, and the
assumption that a single molecular SFL holds across the disk of the
galaxy, all add up to make the systematic uncertainty in the absolute
value of $X_{CO}$ values large. An important assumption made is that
all these uncertain scaling factors do not have a strong radial
dependance, therefore allowing for the relative radial trend in
$X_{CO}$ to be measured. This assumption is supported by the fact that
the six combinations of datasets used in this study, as well as
independent measurements from dust SED modelling (Karin Sandstrom private communication), show similar
trend of an increasing $X_{CO}$ with radius. 

In the future, we expect to extend this study to a larger sample of galaxies in the
VENGA survey, in order to confirm if the observed trends are a common
feature among massive spirals in the local universe. In the near
future, ALMA will allow us to study molecular complexes in nearby
galaxies like NGC 628, in the same level of detail that we can
currently achieve in the MW. In the context of this study, the
commissioning of the MUSE IFU in the Very Large Telescope (VLT), will
provide an instrument with excellent spatial resolution over a large
field-of-view, which could be the perfect complement for ALMA
observations of nearby galaxies.\\

The VENGA collaboration acknowledges the generous support from the
Norman Hackerman Advanced Research Program (NHARP)
ARP-03658-0234-2009, G.A.B. acknowledges the support of Sigma Xi,
The Scientific Research Society, Grant in Aid of
Research. N.J.E. ackowledges the support of NSF grant AST 1109116. A.D.B. acknowledges partial support from grants NSF
AST-0838178, NSF AST-0955836, as well as a Cottrell Scholar award from
the Research Corporation for Science Advancement. N.D. acknowledges
support from PAPIIT grant IA-100212. We also acknowledge Desika
Narayanan for providing the curves presented in Figure \ref{fig-5},
and for useful advice regarding the application of his model, and
Karin Sandstrom useful discussions and for providing her data for comparison.
Support for CARMA construction was derived from the Gordon and Betty
Moore Foundation, the Eileen and Kenneth Norris Foundation, the
Caltech Associates, the states of California, Illinois, and Maryland,
and the NSF. Funding for ongoing CARMA development and operations are
supported by NSF and CARMA partner universities. The construction of
the Mitchell Spectrograph (formerly VIRUS-P) was 
possible thanks to the generous support of the Cynthia \& George
Mitchell Foundation. This research has made use of NASA's Astrophysics
Data System, and of the NASA/IPAC Extragalactic Database (NED) which
is operated by the Jet Propulsion Laboratory, California Institute of
Technology, under contract with the National Aeronautics and Space
Administration.

\end{document}